\documentclass[pre,twocolumn,twoside,byrevtex,superscriptaddress,floatfix]{revtex4-1}

\usepackage [english]{babel}
\usepackage{indentfirst}
\usepackage[section]{placeins}
\usepackage{amssymb}
\usepackage{gensymb}
\usepackage{amsmath}
\usepackage{graphics}
\usepackage{chngcntr}
\usepackage{array}
\usepackage{hyperref}
\hypersetup{
    colorlinks=true,
    linkcolor=black,
    citecolor = black,
    urlcolor=blue,
}
\usepackage{rotating}
\usepackage{placeins}
\usepackage{soul}
\newcolumntype{L}{>{\centering\arraybackslash}m{6cm}}

\usepackage{currfile}
\usepackage{graphicx,epsfig,verbatim,enumerate}
\usepackage{amssymb,amsmath}
\usepackage{ifthen}

\usepackage{longtable}

\usepackage{mathtools}

\newboolean{twocolswitch}

\newcommand{\sindex}[1]{}
\newcommand{\nindex}[1]{}

\newcommand{\www}[1]{\url{#1}}

\usepackage{lettrine}

\usepackage{color}

\setboolean{twocolswitch}{true}

\begin{document}

\title{\protect
Public Opinion Polling with Twitter
}

\author{
\firstname{Emily M.}
\surname{Cody}
}
\email{emily.cody@uvm.edu}
\affiliation{Department of Mathematics \& Statistics,
  Vermont Complex Systems Center,
  Computational Story Lab,
  \& the Vermont Advanced Computing Core,
  The University of Vermont,
  Burlington, VT 05401.}

\author{
\firstname{Andrew J.}
\surname{Reagan}
}
\email{andrew.reagan@uvm.edu}
\affiliation{Department of Mathematics \& Statistics,
  Vermont Complex Systems Center,
  Computational Story Lab,
  \& the Vermont Advanced Computing Core,
  The University of Vermont,
  Burlington, VT 05401.}
  
 \author{
\firstname{Peter Sheridan}
\surname{Dodds}
}
\email{peter.dodds@uvm.edu}
\affiliation{Department of Mathematics \& Statistics,
  Vermont Complex Systems Center,
  Computational Story Lab,
  \& the Vermont Advanced Computing Core,
  The University of Vermont,
  Burlington, VT 05401.}
  
    \author{
\firstname{Christopher M.}
\surname{Danforth}
}
\email{chris.danforth@uvm.edu}
\affiliation{Department of Mathematics \& Statistics,
  Vermont Complex Systems Center,
  Computational Story Lab,
  \& the Vermont Advanced Computing Core,
  The University of Vermont,
  Burlington, VT 05401.}

\date{\today}

\begin{abstract}
  \protect
  Solicited public opinion surveys reach a limited subpopulation of willing participants and are expensive to conduct, leading to poor time resolution and a restricted pool of expert-chosen survey topics.  In this study, we demonstrate that unsolicited public opinion polling through sentiment analysis applied to Twitter correlates well with a range of traditional measures, and has predictive power for issues of global importance. We also examine Twitter's potential to canvas topics seldom surveyed, including ideas, personal feelings, and perceptions of commercial enterprises.  Two of our major observations are that appropriately filtered Twitter sentiment (1) predicts President Obama's job approval three months in advance, and (2) correlates well with surveyed consumer sentiment. To make possible a full examination of our work and to enable others' research, we make public over 10,000 data sets, each a seven-year series of daily word counts for tweets containing a frequently used search term.           
\end{abstract}

\pacs{89.65.-s,89.75.Da,89.75.Fb,89.75.-k}

\maketitle

\section{Introduction}

Public opinion data can be used to determine public awareness, to predict outcomes of events, and to infer characteristics of human behaviors. Indeed, readily available public opinion data is valuable to researchers, policymakers, marketers, and many other groups, but is difficult to generate. Solicited polls can be expensive, prohibitively time consuming, and may only reach a limited number of people on a limited number of days. Polling accuracy evidently relies on accessing representative populations and high response rates. Poor temporal sampling will weaken any poll's value as individual opinions vary in time and in response to social influence\cite{salganik2006experimental,cialdini1987influence}.

With the continued rise of social media as a communication platform, the ability to construct unsolicited public opinion polls has become a possibility for researchers though parsing of massive text-based datasets.  Social media provides extraordinary access to public expressions in real time, and has been shown to play a role in human behavior \cite{glenski2016rating}.  

With its open platform, Twitter has proved to be a boon for many research enterprises \cite{pak2010twitter}, having been used to explore a variety of social and linguistic phenomena \cite{cao2012whisper,lin2013voices,lin2014rising}; harnessed as a data source to create an earthquake reporting system in Japan \cite{sakaki2010earthquake}; made possible detection of influenza outbreaks \cite{aramaki2011twitter};  and used to analyze overall public health \cite{paul2011you}.  Predictions made using Twitter have focused on elections \cite{gayo2013meta,tumasjan2010predicting}, the spread of disease \cite{ritterman2009using}, crime \cite{wang2012automatic}, and the stock market \cite{mittal2012stock}.  These studies demonstrate a proof-of-concept, avoiding the more difficult task of building operational systems for continued forecasting.  

We must be clear that for all its promise, prediction via social media is difficult.  Indeed, we have seen a number of high profile failures such as Google Flu trends \cite{lazer2014a} and various attempts to predict election outcomes \cite{gayo-avello2012a}.  OpinionFinder was used in \cite{bollen2011twitter} to evaluate tweets containing direct expressions of emotion as in `I feel', or `I am feeling', and shown not to have predictive power for the stock market  

Despite limitations which we address later in Sec. IV, Twitter data reveals an unprecedented view of human behavior and opinion related to major issues of global importance \cite{mejova2015twitter}.  In a previous study \cite{cody2015climate}, we analyzed the sentiment surrounding climate change conversation on Twitter.  We discovered that sentiment varies in response to climate change news and events, and that the conversation is dominated by activists.  Another study by Helmuth et al. analyzed tweets by United States Senators to determine which research oriented science organizations and which senators are best at getting science-related findings into the hands of the general public \cite{helmuth2016trust}.  Twitter is also often used to analyze public opinion of political issues \cite{digrazia2013more, barberaless, vaccari2013social}, and in several previous works as an opinion polling resource.  In an application using neural networks called TrueHappiness, users enter one of 300,000 words to obtain a sentiment estimation based on this word's usage in a massive Wikipedia data set, and on previously collected sentiment scores for 10,222 words on Amazon's Mechanical Turk \cite{dodds2011temporal,diehl2016truehappiness}, hereafter referred to as the labMT word set  In another application called RACCOON, users are invited to enter a query term and a rough sketch to obtain words or phrases on Twitter that correlate well with the inputs \cite{raccoon}.  Google Correlate is a similar tool that discovers Google searches for terms or phrases that match well with real-world time series \cite{google}.  Financial term searches from Google Trends was shown by Preis et al. \cite{preis2013quantifying} to correlate with Dow Jones economic indices.    

We argue that Twitter is a better source for opinion mining than Wikipedia, used in TrueHappiness, due to the personal nature of each post.  RACCOON uses only user estimates for correlations and not actual survey data.  Google Correlate uses only frequencies of Google searches to compare time series and may only be useful in specific situations.  

In many studies that use text data from Twitter, the results are not compared to actual polling data, leaving the conclusions open to interpretation.  One example work which does make a direct comparison is \cite{o2010tweets}, where the authors use a Twitter data set from 2008 and 2009 to compare sentiments on Twitter, calculated with OpinionFinder, with daily and monthly public opinion polls.  Our approach here is analogous to that of \cite{o2010tweets}, but we use the sentiment analysis techniques developed in \cite{cody2015climate} to investigate public opinion regarding over 10,000 search terms.  

Specifically, for each of 10,222 of the most frequently used English words, we calculate daily, weekly, monthly, and yearly sentiment time series of co-occurring words from September 2008 to November 2015.  We compare many of these happiness time series to actual polling data, which is not typically available at such a high resolution in time.  We investigate a wide range of topics, including politics, the economy, and several commercial organizations.  Given the size of the dataset, we are unable to exhaustively compare all search terms to solicited opinion polls, and have released the data publicly along with this paper (see \url{http://compstorylab.org/share/papers/cody2016a/index.html}).   

Overall, out aim is to determine the extent to which Twitter can be used to complement traditional public opinion surveys, ideally as a dashboard indicator accompanied by solicited feedback.

\section{Methods}
We implement the ``hedonometer", an instrument designed to calculate a happiness score for a large-scale text, based on the happiness of individual words in the text.  The hedonometer uses previously assessed happiness scores for the labMT word set, which contains the most frequently used English words in four disparate corpora \cite{kloumann2012positivity}.  We choose the hedonometer to obtain lexical coverage of Twitter text and produce meaningful word shift graphs \cite{1512.00531}.

The words were scored in isolation by human subjects in previous work on a scale from 1 (least happy) to 9 (most happy).  We remove neutral and ambiguous words (scores between 4 and 6) from the analysis.  For details regarding stop words, see Dodds et al. \cite{dodds2011temporal}.

We use the hedonometer to calculate what we refer to as \textit{ambient happiness} scores for each of the labMT words (first defined in \cite{dodds2011temporal}).  We determine ambient happiness of a given word, $w_j$, by calculating the average happiness of the words that appear in tweets with that word, i.e., 
\begin{equation}
 h_{\text{amb}}(w_j) = \frac{\sum\limits_{i=1,i\ne j}^{N}h_{\text{avg}}(w_i)f_i}{\sum\limits_{i=1,i\ne j}^{N}f_i}.
 \label{amb}
 \end{equation}
Here, $w_i$ is a word that appears in a tweet with word $w_j$, $h_{\text{avg}}(w_i)$ is the surveyed happiness score of word $i$, $f_i$ is the frequency of word $i$, and $N$ is the number of words in labMT (with stop words removed) appearing in tweets containing word $j$.  Note that we do not include the frequencies or happiness scores of the given word ($w_j$) in the calculation of ambient happiness.

For example, $h_{\text{avg}}=8.42$ for ``love" and the ambient happiness for tweets containing ``love", averaged using Eqn \ref{amb} over the 7 year period, is 6.17.  For ``hate", $h_{\text{avg}}=2.34$ and we find the ambient happiness of ``hate" is 5.75.  As seen in the Appendix in Fig.~\ref{scatter}, we find that due to averaging, ambient happiness covers a smaller range of scores than labMT happiness for individual words.

We use the ambient happiness scores to create time series for each of the words in the labMT word set, and we correlate the happiness time series with polling data at various temporal resolutions.

\subsection{Data}
We collected tweets from Twitter's gardenhose API from September 2008 to November 2015.  During this time period, the volume of tweets grew by three orders of magnitude, but the random percentage of all public tweets fell from 50\% to 10\%.  For each word in the labMT dictionary, e.g. ``Obama", we subsample the gardenhose for all tweets matching the word.  We then tabulate the daily frequencies of labMT words in this term-defined collection of tweets, resulting in temporal counts of the words co-occuring with ``Obama".  For example, the resulting collection of counts for ``Obama" is a  2,628 (days) by 10,222 (words) matrix with entry $(i,j)$ representing the frequency of labMT word $j$ appearing in a tweet containing the term ``Obama" on day $i$.  This collection of counts is posted on the \href{http://compstorylab.org/share/papers/cody2016a/index.html}{online Appendix} for this paper.

\begin{figure}[tp!]
\centerline{\includegraphics[width=0.55\textwidth]{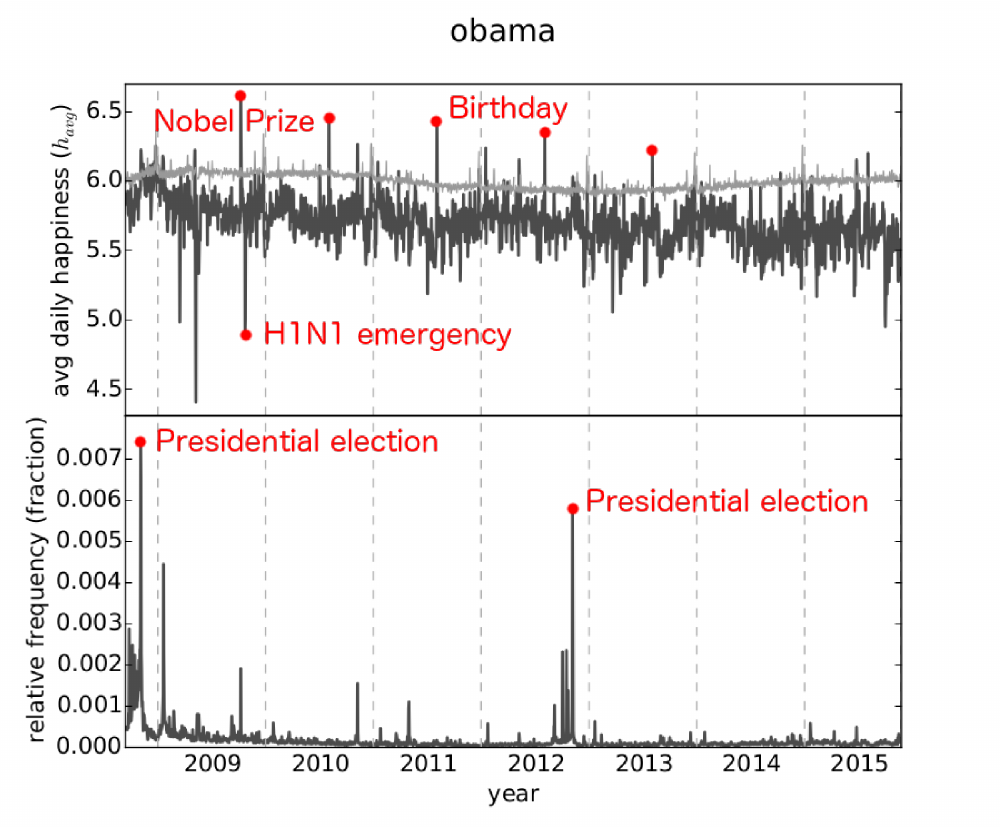}}
\caption{Average daily happiness of tweets containing ``Obama" (top) with the relative frequency of ``Obama" tweets (bottom).  Spikes in happiness include President Obama's birthday (August 4th)  and his winning of the Nobel Prize (10/09/2009).  Dips include a state of emergency for the H1N1 virus.  Spikes in relative frequency occur on election days in 2008 and 2012.}
\label{obamaday}
\end{figure}

Fig.~\ref{obamaday} gives the average daily ambient happiness of  ``Obama", along with the average daily happiness of all tweets during the same time period.  Along with a general slow decline, we see spikes in happiness each year on August 4th, the President's birthday, with the largest spike occurring on October 9, 2009 when President Obama was awarded the Nobel Peace Prize.  We see a strong dip shortly after on October 26, 2009 when President Obama declares a state of emergency for the H1N1 virus.  We see spikes in relative frequency of ``Obama" on both election days in 2008 and in 2012.   

To compare our findings with solicited opinions, we collected yearly and quarterly polling data from Gallup \cite{gallup}.  We focus on quarterly data, as it is the highest resolution we were able to obtain.  The yearly analysis provides us with only 7 data points, and results are in the Appendix.  We compare President Obama's job approval rating on Gallup and on Pollster \cite{pollster}, which allows for daily data collection through their API.  Finally, we use the University of Michigan's Index of Consumer Sentiment data, which is collected monthly \cite{ics}.  

\section{Results}
\subsection{Unsolicited Public Opinions}
\begin{sidewaysfigure*}[!htb]
\centering
\includegraphics[width=1.05\textwidth]{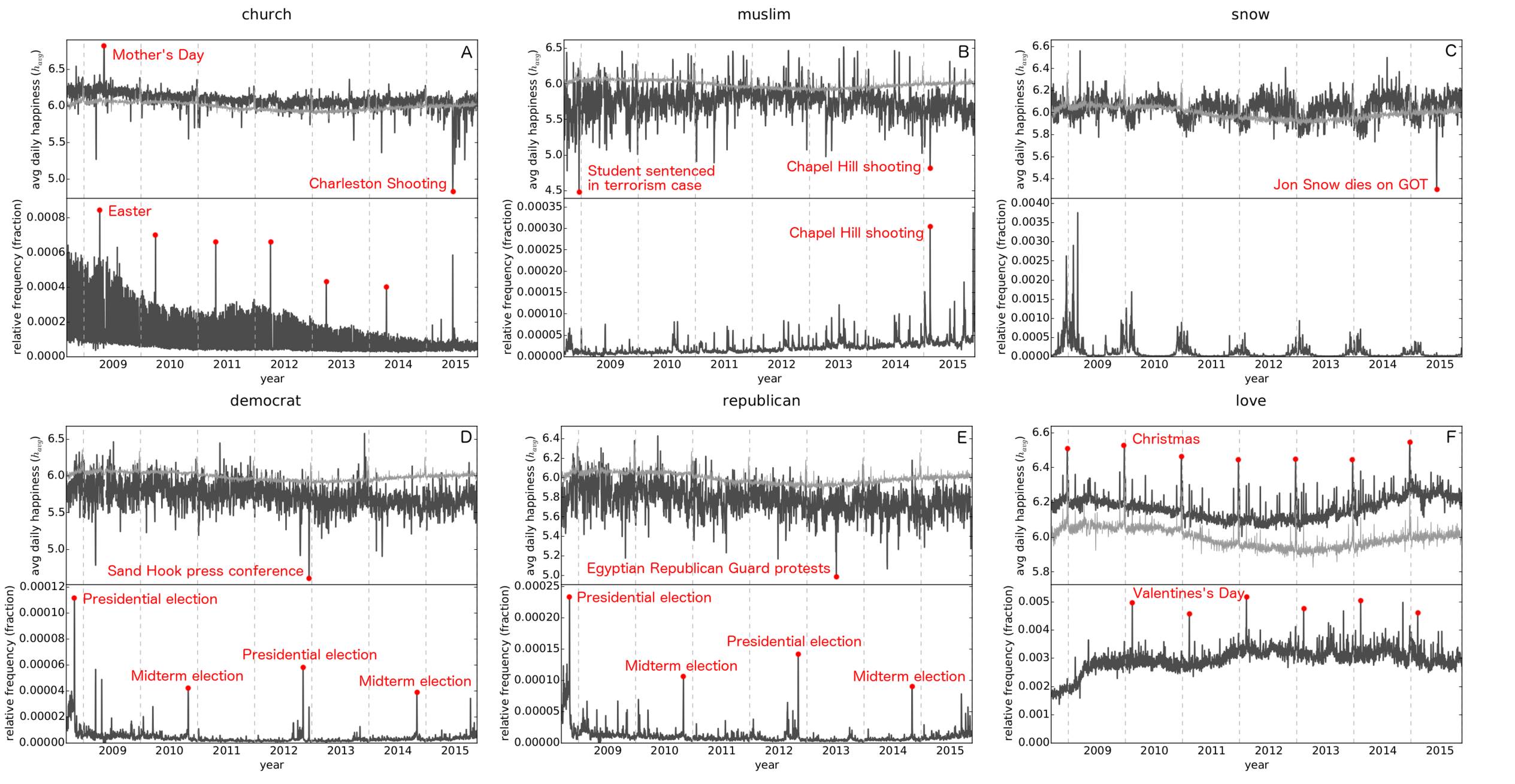}
 \caption{Six examples of ambient happiness time series (top, dark gray) along with relative frequency (bottom).  Twitter's overall average happiness trend is in light gray for each plot.  Relative frequency is approximated by dividing the total frequency of the word by the total frequency of all labMT words on a given day.  (A) ``church": There is a large spike in happiness on Mother's day and a large dip following the Charleston church shooting in June 2015.  There are spikes in relative frequency each Sunday, and yearly on Easter Sunday.  (B) ``muslim": Two dips correspond to a sentencing in a terrorism case in late 2008, and the shooting at Chapel Hill in February 2015. (C) ``snow": Sentiment and relative frequency are seasonal, with a large dip when a main character dies on the HBO show Game of Thrones. (D) ``democrat": Overall sentiment gradually decreases with a large dip after president Obama's press conference following the Sandy Hook shooting.  There are spikes in relative frequency on election days. (E) ``republican": Overall sentiment gradually decreases with a large dip after protests of the Egyptian Republican Guard.  (F) ``love": Sentiment peaks each year on Christmas while relative frequency peaks each year on Valentine's Day.  Weekly and monthly ambient happiness time series for each of these six terms are given in the Appendix (Figs.~\ref{bigfig_week} and \ref{bigfig_month}) and time series for nearly 10,000 terms can be found in the \href{http://compstorylab.org/share/papers/cody2016a/index.html}{online Appendix} for the paper.} 
\label{amb_figs}
\end{sidewaysfigure*}

Here we present happiness time series for several words for which we find interesting patterns.  Figure~\ref{amb_figs} gives examples of ambient happiness and relative frequency time series for a few selected words.  Happiness associated with certain religious words, e.g. ``church" and ``muslim" has decreased in recent years, with dips corresponding to several mass shootings including the Charleston shooting in June 2015 and the Chapel Hill shooting in February 2015.  The relative frequency of ``church" peaks each year on Easter Sunday.  We see that ambient happiness of ``snow'' is seasonal, with the highest happiness during the northern hemisphere summer and lowest during the winter, while the relative frequency is highest during the winter and lowest during the summer.  The saddest day for ``snow'' was June 14, 2015 when a popular Game of Thrones character is presumed dead.  The ambient happiness scores of ``democrat" and ``republican" are on a slow decline, with relative frequencies peaking during presidential and midterm elections.  President Obama's press conference after the Sandy Hook shooting is the saddest day for ``democrat'', while the saddest day for ``republican'' coincides with an incident involving the Egyptian Republican Guard.  Ambient happiness of ``love" peaks around the holidays each year, and the relative frequency was increasing until recently.  While ambient happiness of ``love'' peaks on Christmas each year, the relative frequency of ``love" peaks on Valentine's Day each year, which could be due to the difference in labMT scores for ``christmas'' and ``valentines'' (7.96 and 7.30 respectively).   

In traditional polls, there may be large differences in public opinion from one time period to the next.  In a yes/no or multiple choice survey question it is impossible to use that data to determine why differences occur.  Here we use word shift graphs to determine the cause of a shift in ambient happiness. 

\begin{figure}[tp!]
\centerline{\includegraphics[width=0.44\textwidth]{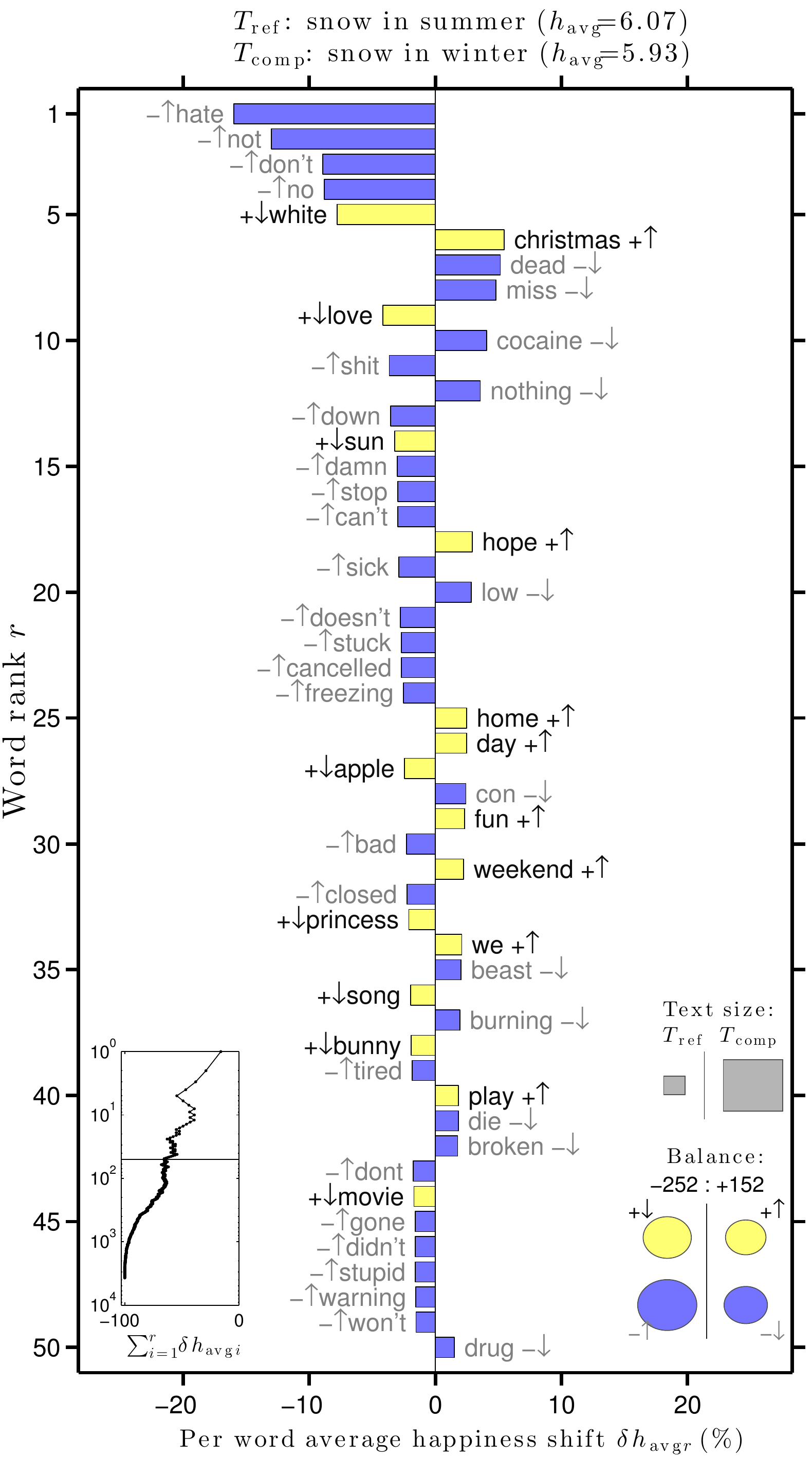}}
\caption{A word shift graph comparing tweets that contain the word ``snow" during the summer months (reference text) and winter months (comparison text).  A purple bar indicates a relatively negative word, a yellow bar indicates a relatively positive word, both with respect to the reference text's average happiness.  An up arrow indicates that word was used more in the comparison text.  A down arrow indicates that word was used less in the comparison text.  Words on the left contribute to a decrease in happiness in the comparison text.  Words on the right contribute to an increase in happiness in the comparison text.  The circles in the lower right corner indicate how many happy words were used more or less and how many sad words were used more or less in the comparison text.}
\label{snowshift}
\end{figure}

A word shift graph ranks words by their contributing factor to the change in happiness between two pieces of text.  For example, in Fig.~\ref{snowshift} we investigate why the ambient happiness of ``snow" is higher in the northern hemisphere summer (when its relative frequency is lowest) and lower in the winter (when its relative frequency is highest). 

The word shift graph in Fig.~\ref{snowshift} compares ``snow" tweets during the winter months (December, January, February) to ``snow" tweets in the summer months (June, July, August).  English speaking countries like Australia and New Zealand will necessarily be included in the wrong season, however their contribution is small. 

 We find that Twitter users loathe the snow during the winter, and miss the snow during the summer, as indicated by the increase in the word ``hate", negatives, and profanity during the winter months and the decrease in the word ``love".  The influence of the Disney classic ``Snow White" is also visible, appearing to be referenced more often in summer months due to its motion picture release on June 1, 2012.    

\FloatBarrier 

\subsection{President Obama's Job Approval Rating}

We next investigate the relationship between President Obama's Job Approval Rating from two public opinion polling resources and the ambient happiness of ``Obama" tweets.     


President Obama's quarterly job approval rating is freely available on \url{gallup.com} \cite{gallup}, and President Obama's daily job approval rating is freely available on \url{pollster.com} \cite{pollster}.  

\begin{figure}[tp!]
\centerline{\includegraphics[width=0.5\textwidth]{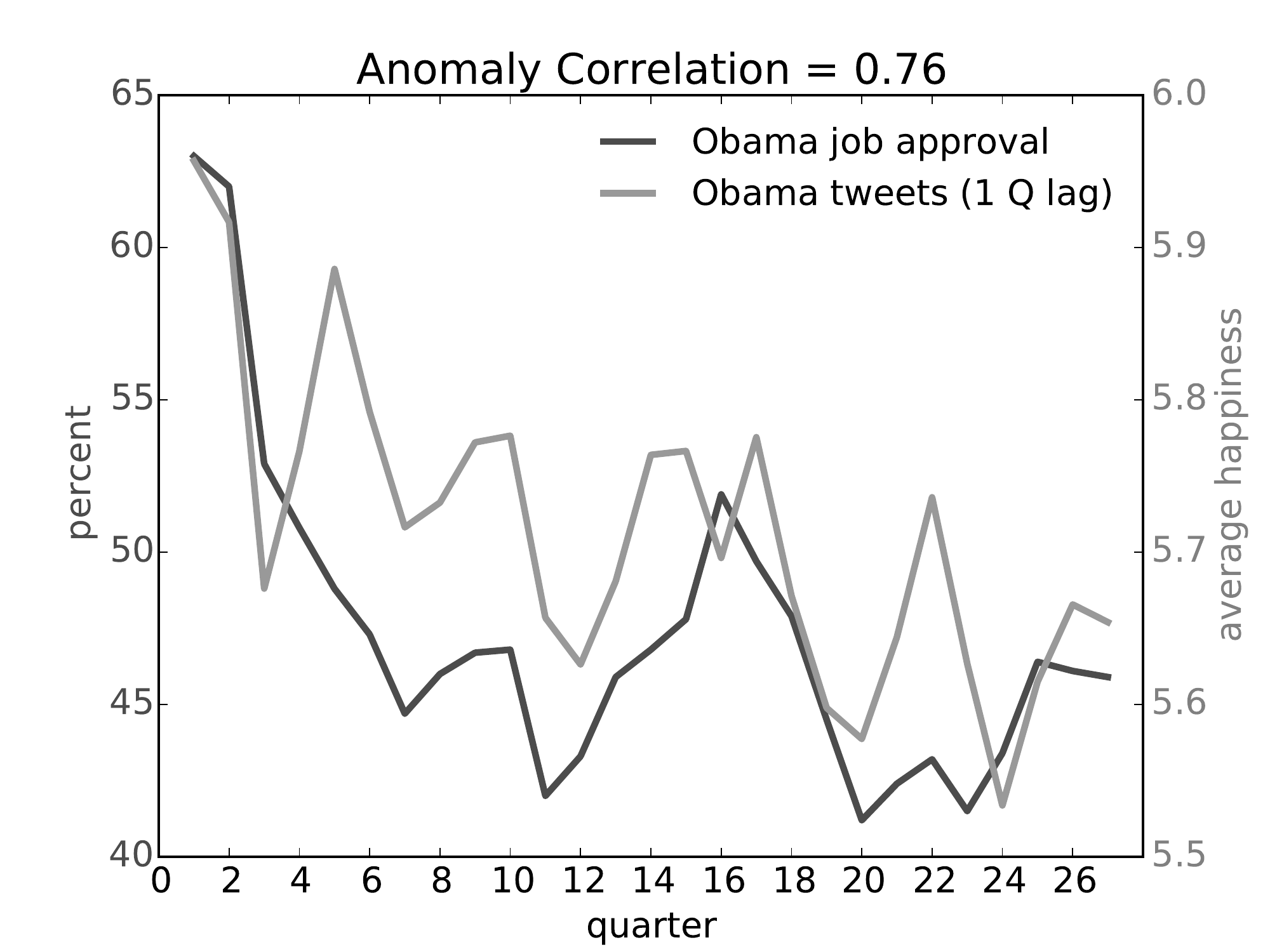}}
\caption{Average quarterly happiness of tweets containing ``Obama" on a one quarter lag with Obama's quarterly job approval rating.  The high positive correlation indicates opinions on Twitter precede timely solicited surveys.}
\label{obamaquarter}
\end{figure}    

We correlate the average quarterly happiness of tweets containing the word ``Obama" with President Obama's quarterly job approval rating and find a strong positive correlation (see Appendix Fig.~\ref{obamaquarterA}).  However, we find the correlation is much stronger in Fig.~\ref{obamaquarter}, which gives the happiness time series at a one quarter lag.  Similarly, we find a strong positive correlation between the daily approval rating available on Pollster and the daily ambient happiness of ``Obama" (see Appendix Fig.~\ref{obamadailyA}a) with an improvement in the correlation when the tweets are lagged by 30 days in Appendix Fig.~\ref{obamadailyA}b.  This indicates that real time Twitter data has the potential to predict solicited public opinion polls.   

\begin{figure}[tp!]
\centerline{\includegraphics[width=0.5\textwidth]{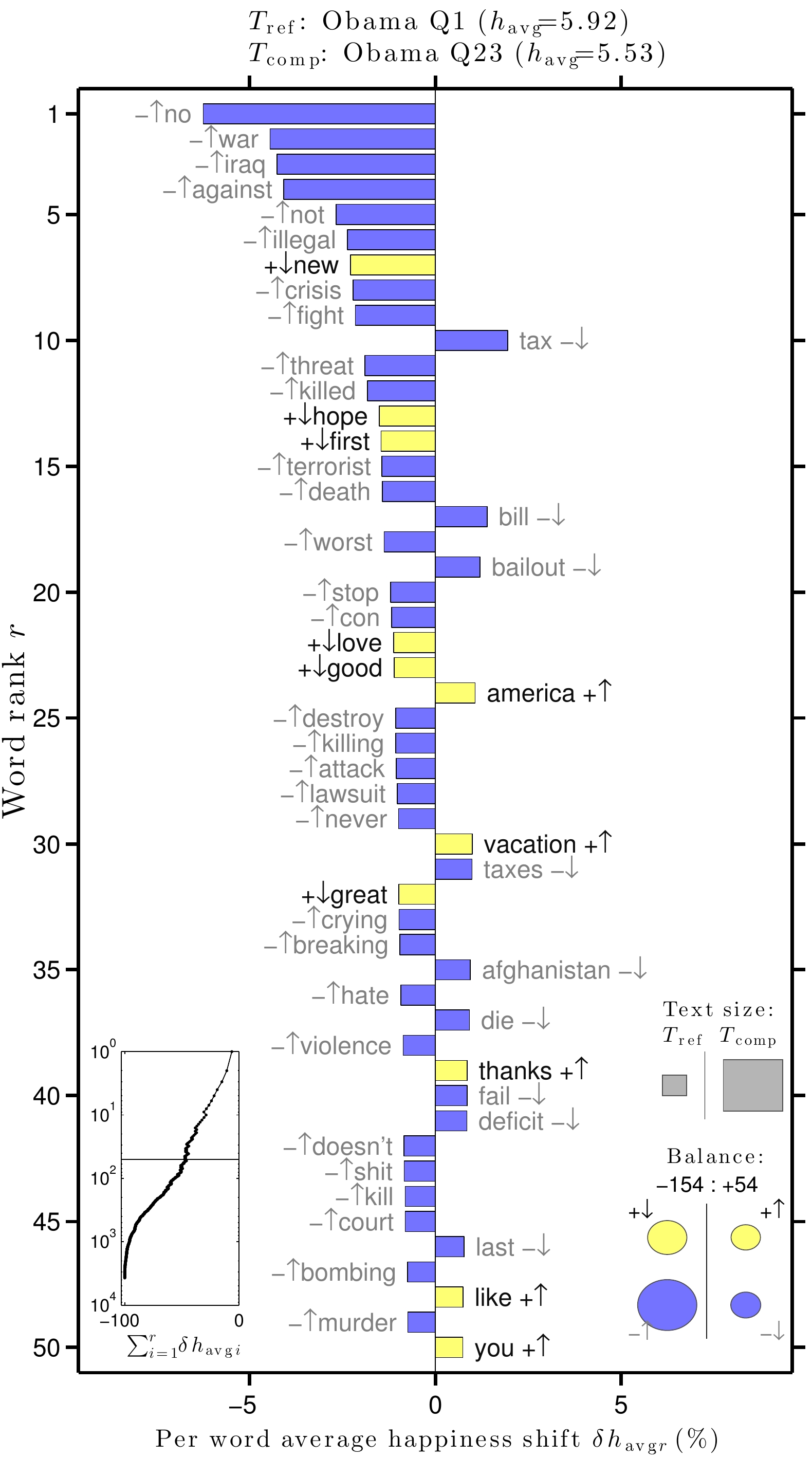}}
\caption{A word shift graph comparing tweets that contain the word ``Obama" during the first quarter of his presidency, 2009/01--2009/03, (reference text) and 23rd quarter of his presidency, 2014/07--2015/09, (comparison text).  Tweets referred to war and terrorism more often in quarter 1.}
\label{obamashift}
\end{figure}

Figure~\ref{obamaquarter} shows that President Obama's highest approval rating in all three sources was during his first quarter (January--March, 2009).  His lowest approval rating was during his 23rd quarter (July--September, 2014).  Fig.~\ref{obamashift} shows which words contributed most to this shift in ambient happiness.  Tweets containing the word ``Obama" discuss war and terrorism more often during his 23rd quarter than his first quarter. 

 \subsection{Index of Consumer Sentiment}
\begin{figure}[tp!]
\centerline{\includegraphics[width=0.4\textwidth]{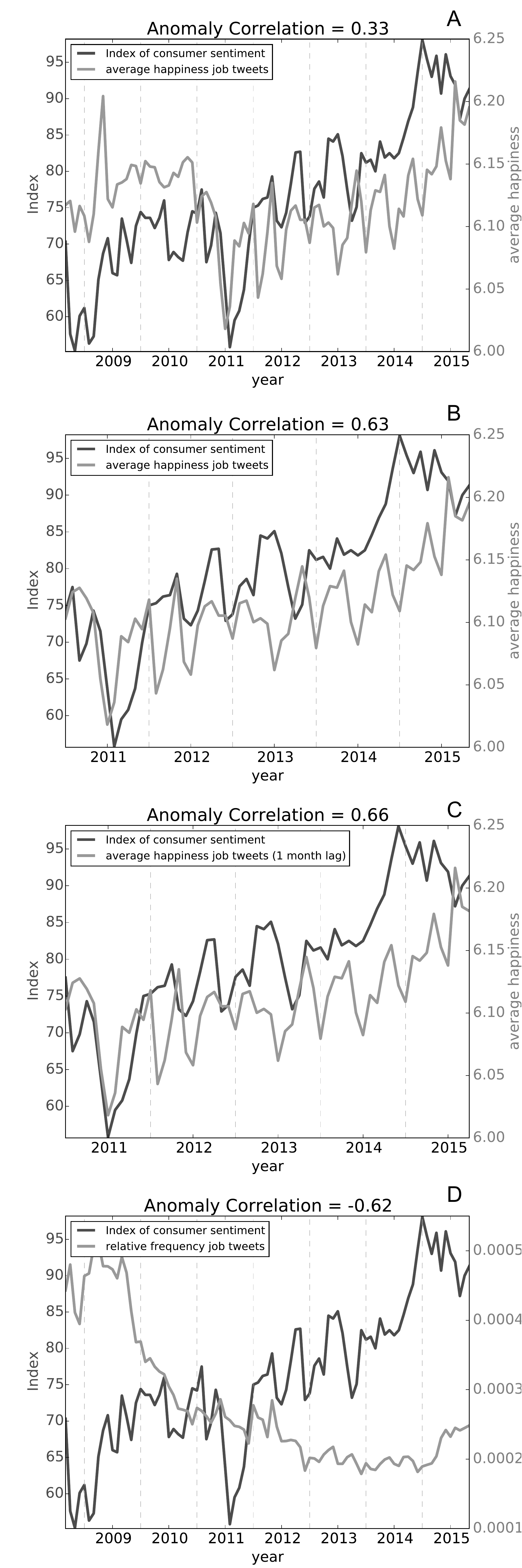}}
\caption{(A) Ambient happiness of ``job" with the Index of Consumer Sentiment.  We see a small positive correlation getting stronger after 2011. (B) Ambient happiness of ``job" with ICS starting in 2011. (C)  Ambient happiness of ``job" is lagged by one month.  (D) ICS with relative frequency of ``job".}
\label{ICS}
\end{figure}

Next, we investigate a monthly poll on consumer sentiment designed by the University of Michigan \cite{ics}.  This poll asks participants five questions about their current and future financial well being and calculates an Index of Consumer Sentiment (ICS) based on responses.  In Fig.~\ref{ICS} we correlate this monthly time series with the ambient happiness of the word ``job".  We find that the correlation is much stronger starting in 2011 (Fig.~\ref{ICS}b), and slightly stronger still when the ambient happiness is lagged one month (Fig.~\ref{ICS}c).  In Fig.~\ref{ICS}d we correlate the ICS with the relative frequency of the word ``job" on Twitter.  We find a strong negative correlation, indicating that it is more likely that a user will tweet the word ``job" when they are searching for one.  

\clearpage

\subsection{Business Sentiment Shifts}
\begin{figure}[tp!]
\centerline{\includegraphics[width=.5\textwidth]{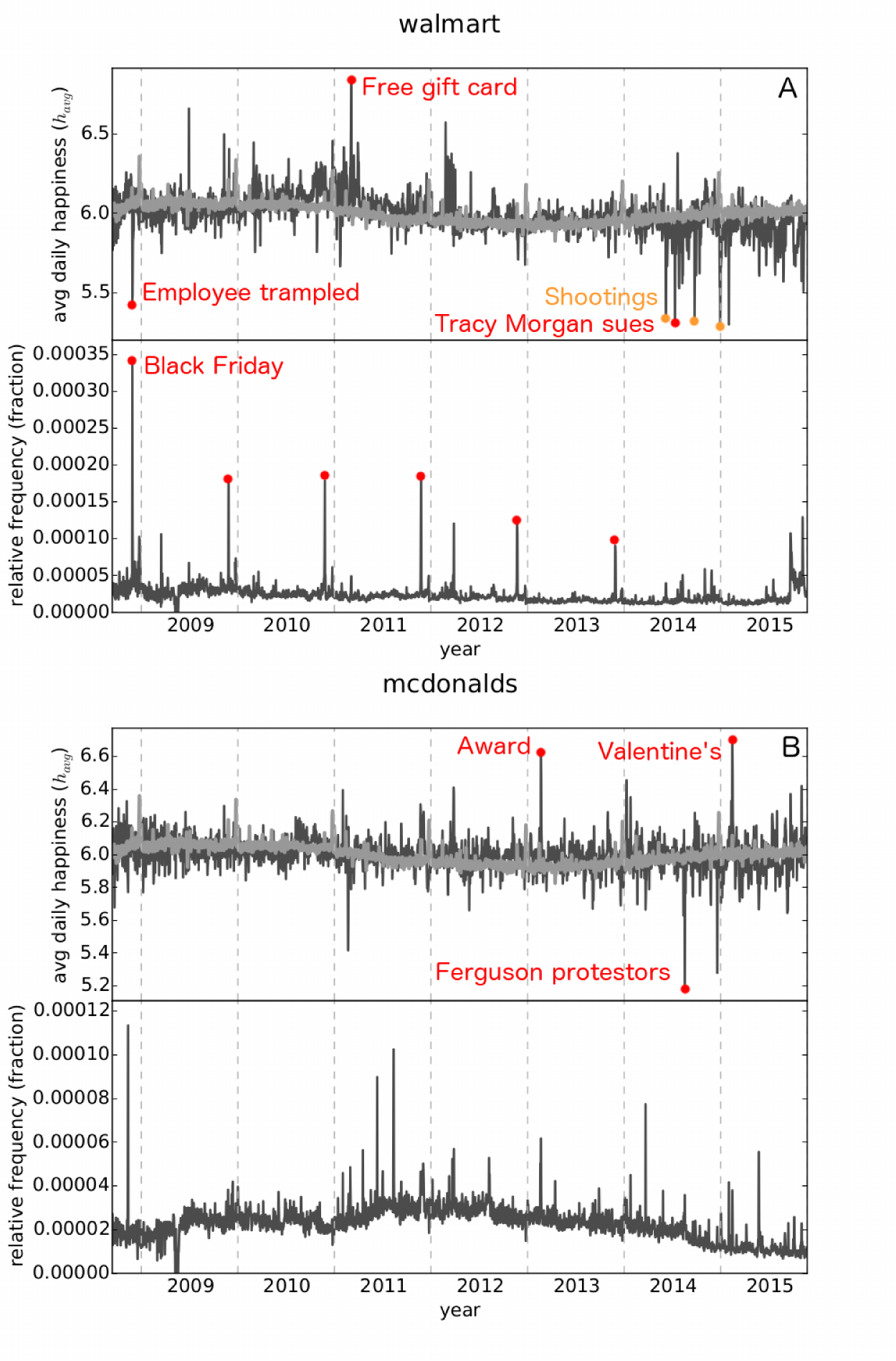}}
\caption{The ambient happiness and relative frequency time series for (A) ``walmart" and  (B) ``mcdonalds''.  Dips in sentiment correspond to deaths, lawsuits, and protests, while spikes in happiness correspond to awards, giveaways, and holidays.  Spikes in the relative frequency of ``walmart" appear largely on Black Friday.  Time series for nearly 10,000 other terms can be found on the \href{http://compstorylab.org/share/papers/cody2016a/index.html}{online Appendix} for the paper.}
\label{business}
\end{figure}

In this section, we investigate the changes in Twitter sentiment surrounding two businesses, Walmart and McDonalds.  We examine the ambient happiness time series to determine how sentiment changes in response to events that took place at specific stores.  Fig.~\ref{business} gives the ambient happiness and relative frequency of the words ``walmart" and ``mcdonalds".     

Many of the spikes in the ``walmart" ambient happiness time series correspond to free giveaways to which Twitter users are responding.  A dip in November 2008 corresponds to the trampling to death of a Walmart employee on Black Friday (the day after Thanksgiving, notorious in the U.S. for shopping).  Shootings that took place in Walmart stores in 2014 are shown with orange dots in Fig.~\ref{business}a.  In June 2014 the Jerad and Amanda Miller Las Vegas shootings ended with 5 deaths (including themselves) in a Nevada Walmart.  In September 2014, the police officer who shot John Crawford in an Ohio Walmart was indicted.  In December 2014, a 2 year old accidentally shot and killed his mother in an Idaho Walmart.  We also see a dip in happiness on the day Tracy Morgan sues Walmart over a fatal crash with one of their tractor trailers in July 2014.   

The happiest day in the ``mcdonalds" ambient happiness time series is Valentine's Day in 2015.  Upon reading some example tweets from this day, we find that McDonalds was a popular ironic destination for Valentine's Day dinner that year among Twitter users.  A second spike corresponds to a prestigious award given to the McDonalds enterprise in February 2013.  McDonalds was given the ``Top Toilet Award" for the cleanliness of its restrooms.  The saddest day for McDonalds on Twitter was August 18, 2014, the day that Ferguson protesters broke into a McDonalds to steal milk to relieve tear gas victims.  

\begin{figure}[tp!]
\centerline{\includegraphics[width=.48\textwidth]{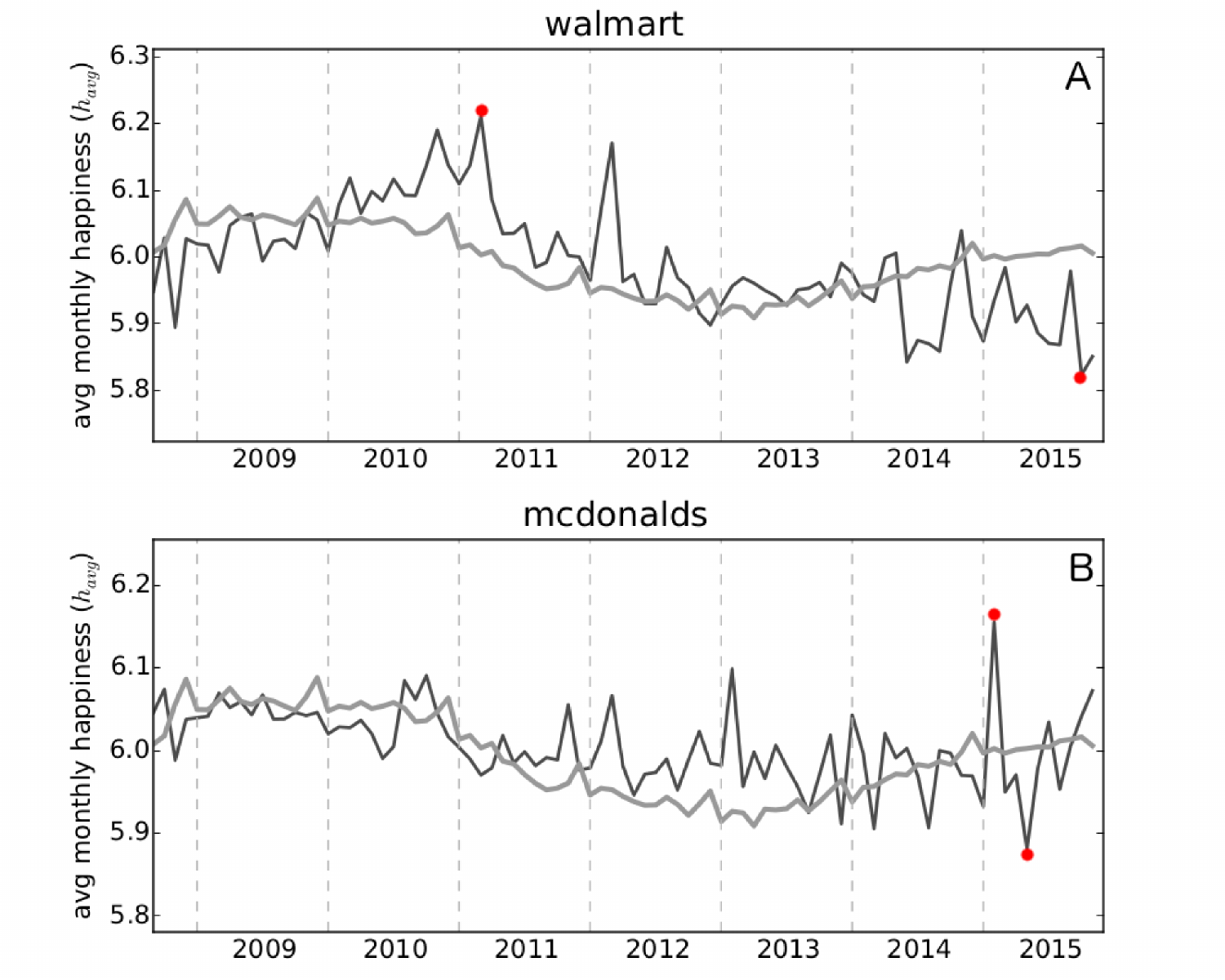}}
\caption{Monthly ambient happiness of (A) ``walmart" and (B) ``mcdonalds".}
\label{bus_by_month}
\end{figure}

\begin{figure*}[h!]
\centerline{\includegraphics[width=\textwidth]{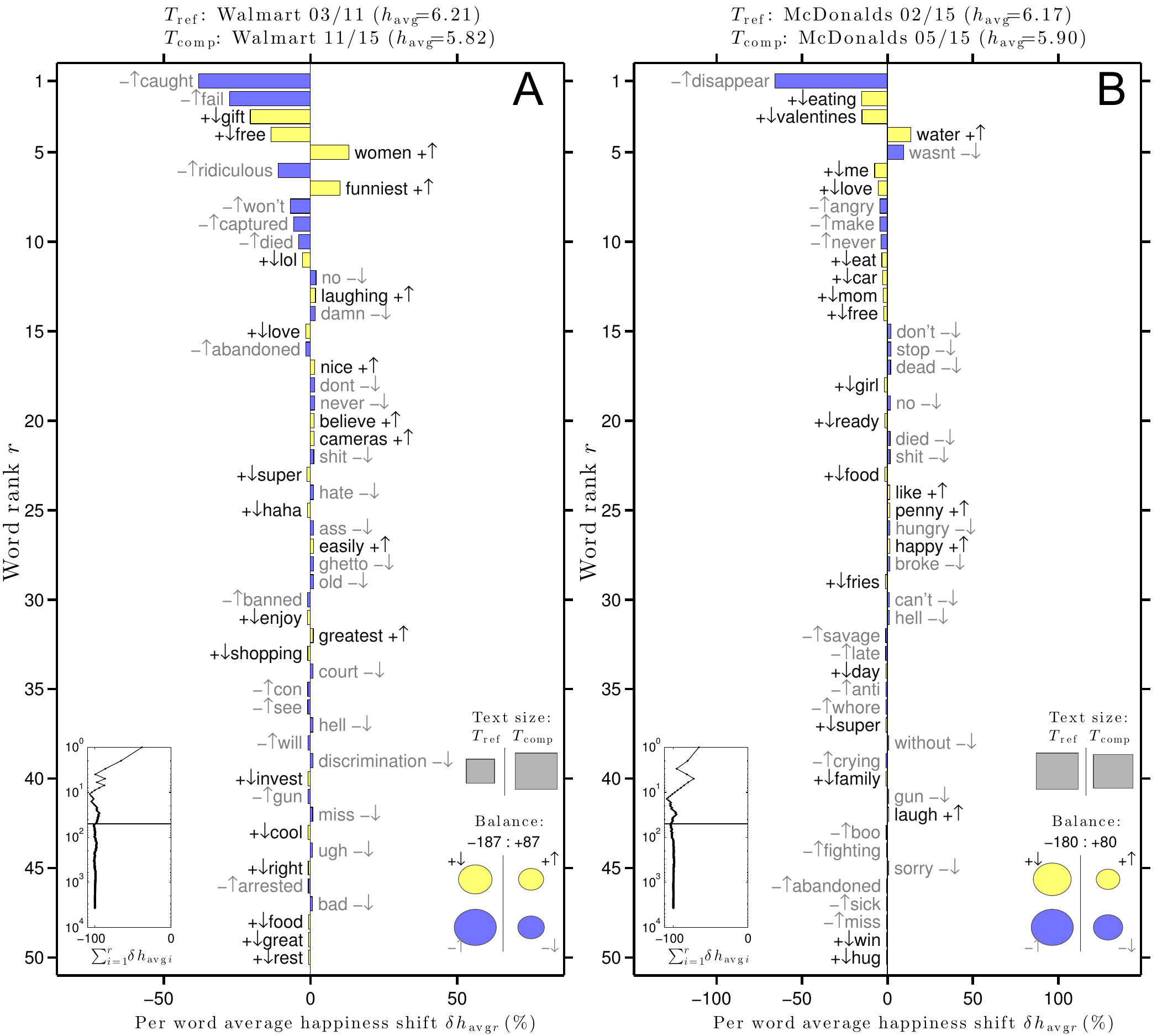}}
\caption{Word shift graphs comparing the happiest and saddest months for (A) ``walmart" and (B) ``mcdonalds".  The happiest month represents the reference text and the saddest month represents the comparison text.}
\label{bus_shifts}
\end{figure*}

In Fig.~\ref{bus_by_month} we explore the monthly ambient happiness of ``walmart" and ``mcdonalds".  We find that the ambient happiness of ``walmart" reaches its maximum in March 2011, and its minimum in October 2015, and the ambient happiness of ``mcdonalds" reaches its maximum in February 2015 and its minimum shortly after in May 2015.  To investigate the texture behind these observations, we use word shift graphs to compare the happiest and saddest months for each business in Fig.~\ref{bus_shifts}.

In November 2015 (comparison text Fig.~\ref{bus_shifts}a), there were Black Friday altercations at many Walmarts throughout the country, often caught on camera, leading to an increase in negative words such as ``caught", ``fail", ``ridiculous", and ``captured".  Twitter users were happier about Walmart in March 2011 (reference text Fig.~\ref{bus_shifts}a) due in part to a free gift card giveaway.  Happier tweets included the words ``lol", ``love",  ``haha", and ``super".  Surprisingly, we actually see more curse words during the happiest month than the saddest month. 

The happiest month for McDonalds was February 2015 (reference text Fig.~\ref{bus_shifts}b) when a surprising number of Twitter users were spending Valentine's Day there, hence the decrease in the words ``valentines" and ``love".  The decrease in happiness in May 2015 is in large part in an increase in the word ``disappear".  During this time, a video of a Michigan McDonalds employee performing a practical joke, in which he claims he's going to make a penny disappear in a bottle of water, went viral.  Using word shifts, we are thus able to determine ``disappear'' was not a true indicator for negativity.  The next step would be to add ``disappear'' to our stop word list and reevaluate the time series.  In general, word shifts are very sensitive diagnostics that allow us to make sense of apparent sentiment patterns, and to adjust the hedonometer as needed.  

\section{Limitations}

Here we present positive correlations between ambient happiness on Twitter and solicited public opinion surveys.  The methods used in this work present several limitations that we address here.  First, we are unable to compare ambient happiness to public opinion surveys for the majority of subjects discussed on Twitter.  Ambient happiness surrounding commercial businesses can be compared to stock prices, however the relationship is unlikely to be linear. Stock prices take into account much more than the success of one business, and thus reflect a different signal than public opinion.

We acknowledge that our dataset can be influenced by small events that are retweeted many times, and cause a transient spike or a dip in ambient happiness. Tweets reflect an broad spectrum of conversation and reactions to major events, including bots and sarcasm to name a few difficulties. The stochastic nature of Twitter timeseries data also makes it difficult to compare high resolution ambient happiness to low resolution opinion surveys.  Appendix  Fig.~\ref{good} shows our attempt to compare ambient happiness to six important topics available on Gallup. Gallup surveys take place over several days, once a year, leading to a natural comparison with ambient happiness averaged for a full 365 days. We are therefore unable to report which method is a more accurate indicator of public opinion. Instead, we conclude that public opinion polling with Twitter has the potential to complement traditional public opinion surveys.

Finally, without knowing user demographics in detail, we are unable to adjust for the known sample bias of Twitter users relative to the general population \cite{pew}.  Our conclusions pertain directly only to the Twitterverse and as suggestive results for the broader public. Twitter users are a mixture of individuals, news outlets, corporations, and, perhaps most problematically, are not even always human. Previous work shows that Twitter contains many bots, which send tweets automatically, often to advertise a product or a political campaign \cite{howard2016bots}.  We do not eliminate these tweets in this work, however many methods for uncovering them have been suggested \cite{ferrara2014rise,chu2012detecting,dickerson2014using,clark2015sifting}.    

\section{Conclusion}
The objective of this research was to determine the extent to which ambient happiness on Twitter can be used as a reliable public opinion polling device.  A central motivation is that solicited public opinion polling data is difficult to obtain at a high temporal resolution, if at all \cite{miller2011social}.

With data from Twitter we can investigate topics other than political or global issues, which are the focus of a large majority of solicited surveys.  We can use ambient happiness to determine how people feel about seemly neutral topics like ``snow", or how they are using the words ``love", and ``feel".  We also have shown that Twitter users respond to various kinds of events taking place at commercial businesses, and thus ambient happiness could be used in market analysis to predict or improve a company's sales.  

Of the available polling data we were able to obtain, we find that ambient happiness of selected words correlates well with solicited public opinions.  In some cases, the correlation increases when the tweets are lagged, indicating that real time Twitter data has the potential to predict solicited public opinion polls.

Not only can tweets anticipate survey responses, but we can use individual words within tweets to determine why one time period is happier than another, something that is not possible in traditional polls due to the multiple choice aspect of most surveys.  Several other advantages of using tweets for public opinion polling include the ability to track movement \cite{frank2013happiness} and make maps \cite{mitchell2013geography, kryvasheyeu2016rapid} using geolocated tweets.  Data collection itself is largely algorithmic, and does not rely on the responses of participants.   


We find that for many topics, Twitter is a valuable resource for mining public opinions without solicited surveys.  We encourage readers to explore the data online at \url{http://compstorylab.org/share/papers/cody2016a/data.html}.  Social media may be the future of public opinion polling, revealing important signals complementary to traditional surveys.  





\acknowledgments
The authors are grateful for the computational resources provided by the Vermont Advanced Computing Core and the Vermont Complex Systems Center.  PSD and CMD acknowledge support from NSF grant \#1447634.  EMC acknowledges support from the National Science Foundation under project DGE-1144388.

\bibliographystyle{abbrv}


\begin{thebibliography}{10}

\bibitem{pew}
The demographics of social media users.
\newblock
  \url{http://www.pewinternet.org/2015/08/19/the-demographics-of-social-media-users/}.
\newblock Accessed: 2016-07-18.

\bibitem{gallup}
Gallup trends.
\newblock \url{http://www.gallup.com/poll/trends.aspx}.
\newblock Accessed: 2016-03-08.

\bibitem{google}
Google correlate.
\newblock \url{https://www.google.com/trends/correlate},.

\bibitem{pollster}
Pollster {API}.
\newblock \url{http://elections.huffingtonpost.com/pollster/api}.
\newblock Accessed: 2016-03-08.

\bibitem{ics}
University of {M}ichigan {I}ndex of {C}onsumer {S}entiment.
\newblock \url{http://www.sca.isr.umich.edu/tables.html}.
\newblock Accessed: 2016-03-08.

\bibitem{raccoon}
D.~Antenucci, M.~R. Andwerson, P.~Zhao, and M.~Cafaerlla.
\newblock A query system for social media signals.
\newblock 2015.

\bibitem{aramaki2011twitter}
E.~Aramaki, S.~Maskawa, and M.~Morita.
\newblock {T}witter catches the flu: {D}etecting influenza epidemics using
  {T}witter.
\newblock In {\em Proceedings of the conference on empirical methods in natural
  language processing}, pages 1568--1576. Association for Computational
  Linguistics, 2011.

\bibitem{barberaless}
P.~Barber{\'a}.
\newblock Less is more? {H}ow demographic sample weights can improve public
  opinion estimates based on {T}witter data.

\bibitem{bollen2011twitter}
J.~Bollen, H.~Mao, and X.~Zeng.
\newblock Twitter mood predicts the stock market.
\newblock {\em Journal of Computational Science}, 2(1):1--8, 2011.

\bibitem{cao2012whisper}
N.~Cao, Y.-R. Lin, X.~Sun, D.~Lazer, S.~Liu, and H.~Qu.
\newblock Whisper: {T}racing the spatiotemporal process of information
  diffusion in real time.
\newblock {\em Visualization and Computer Graphics, IEEE Transactions on},
  18(12):2649--2658, 2012.

\bibitem{chu2012detecting}
Z.~Chu, S.~Gianvecchio, H.~Wang, and S.~Jajodia.
\newblock Detecting automation of {T}witter accounts: {A}re you a human, bot,
  or cyborg?
\newblock {\em Dependable and Secure Computing, IEEE Transactions on},
  9(6):811--824, 2012.

\bibitem{cialdini1987influence}
R.~B. Cialdini and N.~Garde.
\newblock {\em Influence}, volume~3.
\newblock A. Michel, 1987.

\bibitem{clark2015sifting}
E.~M. Clark, J.~R. Williams, C.~A. Jones, R.~A. Galbraith, C.~M. Danforth, and
  P.~S. Dodds.
\newblock Sifting robotic from organic text: {A} natural language approach for
  detecting automation on {T}witter.
\newblock {\em Journal of Computational Science}, 2015.

\bibitem{cody2015climate}
E.~M. Cody, A.~J. Reagan, L.~Mitchell, P.~S. Dodds, and C.~M. Danforth.
\newblock Climate change sentiment on {T}witter: {A}n unsolicited public
  opinion poll.
\newblock {\em {P}{L}o{S} ONE}, 10(8):e0136092, 2015.

\bibitem{dickerson2014using}
J.~P. Dickerson, V.~Kagan, and V.~Subrahmanian.
\newblock Using sentiment to detect bots on {T}witter: {A}re humans more
  opinionated than bots?
\newblock In {\em Advances in Social Networks Analysis and Mining (ASONAM),
  2014 IEEE/ACM International Conference on}, pages 620--627. IEEE, 2014.

\bibitem{diehl2016truehappiness}
P.~U. Diehl, B.~U. Pedroni, A.~Cassidy, P.~Merolla, E.~Neftci, and G.~Zarrella.
\newblock True{H}appiness: {N}euromorphic emotion recognition on {T}rue{N}orth.
\newblock {\em arXiv preprint arXiv:1601.04183}, 2016.

\bibitem{digrazia2013more}
J.~DiGrazia, K.~McKelvey, J.~Bollen, and F.~Rojas.
\newblock More tweets, more votes: {S}ocial media as a quantitative indicator
  of political behavior.
\newblock {\em PLoS ONE}, 8(11):e79449, 2013.

\bibitem{dodds2011temporal}
P.~S. Dodds, K.~D. Harris, I.~M. Kloumann, C.~A. Bliss, and C.~M. Danforth.
\newblock Temporal patterns of happiness and information in a global social
  network: {H}edonometrics and {T}witter.
\newblock {\em PloS one}, 6(12):e26752, 2011.

\bibitem{ferrara2014rise}
E.~Ferrara, O.~Varol, C.~Davis, F.~Menczer, and A.~Flammini.
\newblock The rise of social bots.
\newblock {\em arXiv preprint arXiv:1407.5225}, 2014.

\bibitem{frank2013happiness}
M.~R. Frank, L.~Mitchell, P.~S. Dodds, and C.~M. Danforth.
\newblock Happiness and the patterns of life: {A} study of geolocated tweets.
\newblock {\em Scientific reports}, 3, 2013.

\bibitem{gayo-avello2012a}
D.~Gayo-Avello.
\newblock "i wanted to predict elections with twitter and all i got was this
  lousy paper"---a balanced survey on election prediction using twitter data.
\newblock {\em arXiv preprint arXiv:1204.6441}, 2012.

\bibitem{gayo2013meta}
D.~Gayo-Avello.
\newblock A meta-analysis of state-of-the-art electoral prediction from
  {T}witter data.
\newblock {\em Social Science Computer Review}, page 0894439313493979, 2013.

\bibitem{glenski2016rating}
M.~Glenski and T.~Weninger.
\newblock Rating effects on social news posts and comments.
\newblock {\em arXiv preprint arXiv:1606.06140}, 2016.

\bibitem{helmuth2016trust}
B.~Helmuth, T.~C. Gouhier, S.~Scyphers, and J.~Mocarski.
\newblock Trust, tribalism and tweets: {H}as political polarization made
  science a ``wedge issue''?
\newblock {\em Climate Change Responses}, 3(1):1, 2016.

\bibitem{howard2016bots}
P.~N. Howard and B.~Kollanyi.
\newblock Bots,\# {S}trongerin, and\# {B}rexit: {C}omputational propaganda
  during the {U}{K}-{E}{U} referendum.
\newblock {\em Available at SSRN 2798311}, 2016.

\bibitem{kamvar2009we}
S.~Kamvar and J.~Harris.
\newblock {\em We feel fine: {A}n almanac of human emotion}.
\newblock Simon and Schuster, 2009.

\bibitem{kloumann2012positivity}
I.~M. Kloumann, C.~M. Danforth, K.~D. Harris, C.~A. Bliss, and P.~S. Dodds.
\newblock Positivity of the {E}nglish language.
\newblock {\em PLoS ONE}, 7(1):e29484, 2012.

\bibitem{kryvasheyeu2016rapid}
Y.~Kryvasheyeu, H.~Chen, N.~Obradovich, E.~Moro, P.~Van~Hentenryck, J.~Fowler,
  and M.~Cebrian.
\newblock Rapid assessment of disaster damage using social media activity.
\newblock {\em Science Advances}, 2(3):e1500779, 2016.

\bibitem{lazer2014a}
D.~Lazer, R.~Kennedy, G.~King, and A.~Vespignani.
\newblock The parable of {G}oogle {F}lu: {T}raps in {B}ig {D}ata analysis.
\newblock {\em Science Magazine}, 343:1203--1205, 2014.

\bibitem{lin2014rising}
Y.-R. Lin, B.~Keegan, D.~Margolin, and D.~Lazer.
\newblock Rising tides or rising stars?: {D}ynamics of shared attention on
  {T}witter during media events.
\newblock {\em PloS one}, 9(5):e94093, 2014.

\bibitem{lin2013voices}
Y.-R. Lin, D.~Margolin, B.~Keegan, and D.~Lazer.
\newblock Voices of victory: {A} computational focus group framework for
  tracking opinion shift in real time.
\newblock In {\em Proceedings of the 22nd international conference on World
  Wide Web}, pages 737--748. International World Wide Web Conferences Steering
  Committee, 2013.

\bibitem{mejova2015twitter}
Y.~Mejova, I.~Weber, and M.~W. Macy.
\newblock {\em Twitter: {A} digital socioscope}.
\newblock Cambridge University Press, 2015.

\bibitem{miller2011social}
G.~Miller.
\newblock Social scientists wade into the tweet stream.
\newblock {\em Science}, 333(6051):1814--1815, 2011.

\bibitem{mitchell2013geography}
L.~Mitchell, M.~R. Frank, K.~D. Harris, P.~S. Dodds, and C.~M. Danforth.
\newblock The geography of happiness: {C}onnecting {T}witter sentiment and
  expression, demographics, and objective characteristics of place.
\newblock {\em PLoS ONE}, 8(5):e64417, 2013.

\bibitem{mittal2012stock}
A.~Mittal and A.~Goel.
\newblock Stock prediction using {T}witter sentiment analysis.
\newblock {\em Standford University, CS229 (2011 http://cs229. stanford.
  edu/proj2011/GoelMittal-StockMarketPredictionUsingTwitterSentimentAnalysis.
  pdf)}, 2012.

\bibitem{o2010tweets}
B.~O'Connor, R.~Balasubramanyan, B.~R. Routledge, and N.~A. Smith.
\newblock From tweets to polls: {L}inking text sentiment to public opinion time
  series.
\newblock {\em ICWSM}, 11(122-129):1--2, 2010.

\bibitem{pak2010twitter}
A.~Pak and P.~Paroubek.
\newblock Twitter as a corpus for sentiment analysis and opinion mining.
\newblock In {\em LREc}, volume~10, pages 1320--1326, 2010.

\bibitem{paul2011you}
M.~J. Paul and M.~Dredze.
\newblock You are what you tweet: {A}nalyzing {T}witter for public health.
\newblock {\em ICWSM}, 20:265--272, 2011.

\bibitem{preis2013quantifying}
T.~Preis, H.~S. Moat, and H.~E. Stanley.
\newblock Quantifying trading behavior in financial markets using {G}oogle
  {T}rends.
\newblock {\em Scientific reports}, 3, 2013.

\bibitem{1512.00531}
A.~Reagan, B.~Tivnan, J.~R. Williams, C.~M. Danforth, and P.~S. Dodds.
\newblock Benchmarking sentiment analysis methods for large-scale texts: {A}
  case for using continuum-scored words and word shift graphs, 2015.

\bibitem{ritterman2009using}
J.~Ritterman, M.~Osborne, and E.~Klein.
\newblock Using prediction markets and {T}witter to predict a swine flu
  pandemic.
\newblock In {\em 1st international workshop on mining social media}, volume~9,
  pages 9--17. ac. uk/miles/papers/swine09. pdf (accessed 26 August 2015),
  2009.

\bibitem{sakaki2010earthquake}
T.~Sakaki, M.~Okazaki, and Y.~Matsuo.
\newblock Earthquake shakes {T}witter users: {R}eal-time event detection by
  social sensors.
\newblock In {\em Proceedings of the 19th international conference on World
  wide web}, pages 851--860. ACM, 2010.

\bibitem{salganik2006experimental}
M.~J. Salganik, P.~S. Dodds, and D.~J. Watts.
\newblock Experimental study of inequality and unpredictability in an
  artificial cultural market.
\newblock {\em Science}, 311(5762):854--856, 2006.

\bibitem{tumasjan2010predicting}
A.~Tumasjan, T.~O. Sprenger, P.~G. Sandner, and I.~M. Welpe.
\newblock Predicting elections with {T}witter: {W}hat 140 characters reveal
  about political sentiment.
\newblock {\em ICWSM}, 10:178--185, 2010.

\bibitem{vaccari2013social}
C.~Vaccari, A.~Valeriani, P.~Barber{\'a}, R.~Bonneau, J.~T. Jost, J.~Nagler,
  and J.~Tucker.
\newblock Social media and political communication: {A} survey of {T}witter
  users during the 2013 {I}talian general election.
\newblock {\em Rivista italiana di scienza politica}, 43(3):381--410, 2013.

\bibitem{wang2012automatic}
X.~Wang, M.~S. Gerber, and D.~E. Brown.
\newblock Automatic crime prediction using events extracted from {T}witter
  posts.
\newblock In {\em Social Computing, Behavioral-Cultural Modeling and
  Prediction}, pages 231--238. Springer, 2012.

\end{thebibliography}
\clearpage

\appendix
\counterwithin{figure}{section}
\counterwithin{table}{section}
\section{Anomaly Correlation}
We use anomaly correlation (Pearson Correlation) to determine the relationship between the Twitter happiness time series and the polling data.  When doing so, we subtract the mean of the series, $m$, from each data point, $h_i$, to determine anomalies, and then calculate the cosine of the angle between the series of anomalies, i.e.,  
\begin{equation}
H_{an} =  \left\{h_i-m\right\}_{i=1}^L
\end{equation}
\begin{equation}
Corr_{\text{an}}(H,P) =  \frac{H_{an}\cdot P_{an}}{||H_{an}||\cdot||P_{an}||}
\end{equation}
The variables $H$ and $P$ represent happiness time series and polling time series respectively, and $L$ is the length of the time series.  

\section{Additional Figures and Tables}

\begin{figure}[h!]
\centerline{\includegraphics[width=0.5\textwidth]{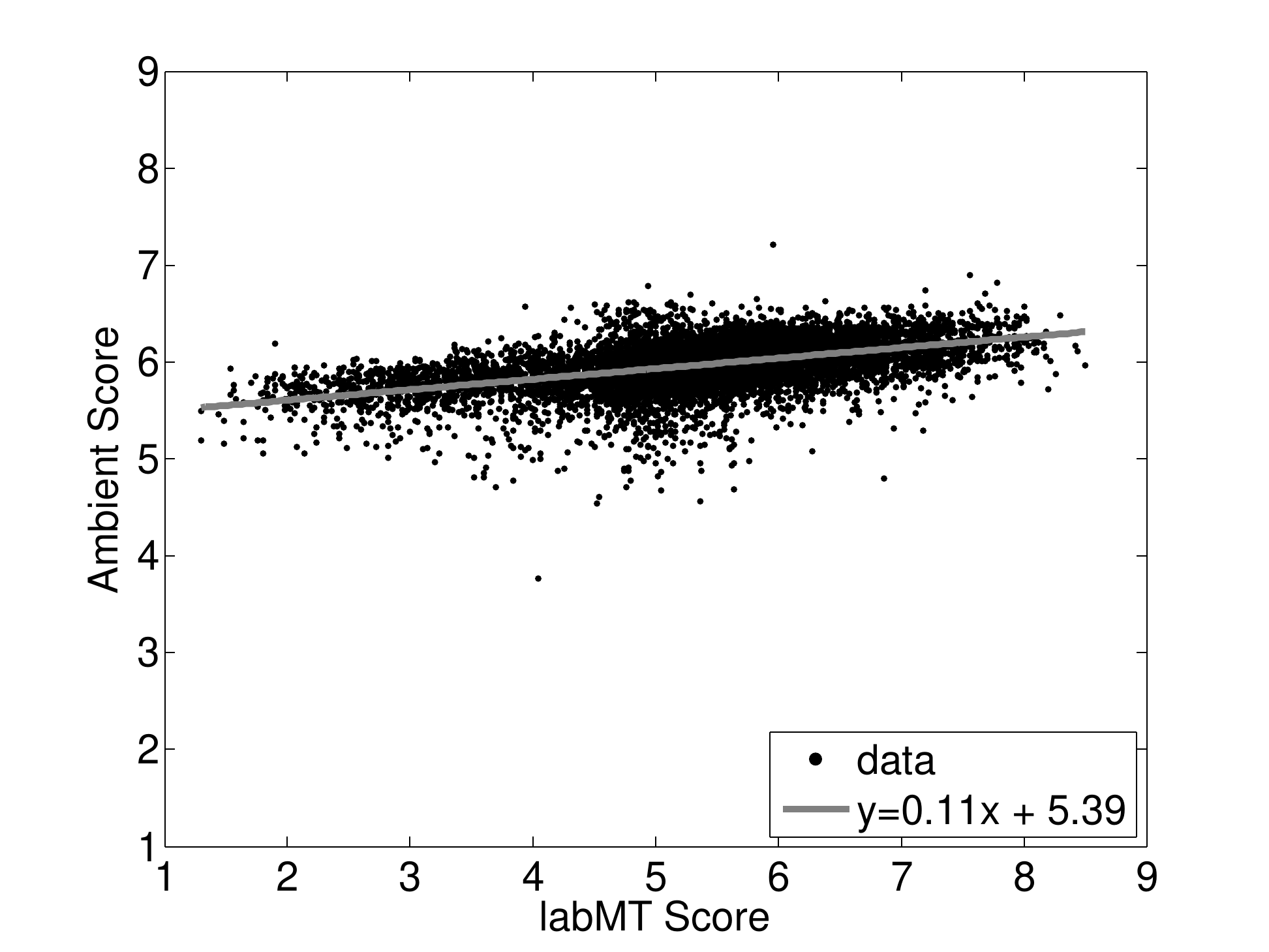}}
\caption{Surveyed happiness versus ambient happiness for all words in the labMT dataset.  The small positive slope indicates that ambient happiness increases with surveyed happiness, however ambient happiness covers a smaller range of values.  An interactive version is available in the online Appendix.}
\label{scatter}
\end{figure}

Each word in our data set was previously assigned a happiness score through Amazon's Mechanical Turk (labMT scores).  We investigate the relationship between surveyed scores and ambient happiness scores in Fig.~\ref{scatter}.  We find a positive slope, indicating that ambient happiness rises with surveyed happiness, however we see a much smaller range of scores, which can be attributed to averaging a large amount of words.  We give the top 10 and bottom 10 words sorted by ambient happiness in Table~\ref{sort_amb}.  Top words included birthday wishes and prize giveaways, and bottom words suggest legal news stories.   

\begin{table}[h]
\resizebox{\columnwidth}{!}{%
\begin{tabular}{lcccclccc}
\\\hline\noalign{\smallskip}
\multicolumn{4}{ c } {\textbf{Top 10}} & \hspace{3mm} & \multicolumn{4}{ c } {\textbf{Bottom 10}}\\\hline\noalign{\smallskip}
Rank & Word & Ambient & labMT && Rank & Word & Ambient & labMT \\\noalign{\smallskip}\hline\noalign{\smallskip}
1. & collected & 7.21 & 5.96 && 9780. & defendants & 4.87 & 4.26\\
2. & merry & 6.90 & 7.56 &&  9781. & prosecutors & 4.87 & 4.20\\
3. & birthday & 6.82 & 7.78 &&  9782. & suspects & 4.86 & 3.60\\
4. & iya & 6.79 & 4.94 && 9783. & suspected & 4.81 & 3.52\\
5. & prizes & 6.73 & 7.20 && 9784. & indicted & 4.81 & 3.60\\
6. & b-day & 6.71 & 7.68 && 9785. & seas & 4.80 & 6.84\\
7. & 2-bath & 6.69 & 5.28 && 9786. & pleaded & 4.78 & 3.84\\
8. & entered & 6.65 & 5.82 && 9787. & sentenced & 4.71 & 3.70\\
9. & giveaway & 6.62 & 6.38 && 9788. & civilians & 4.68 & 5.84\\
10. & shipping & 6.61 & 5.46 && 9789. & welt & 3.77 & 4.04 \\\noalign{\smallskip}\hline
\end{tabular}
}
\caption{The top 10 and bottom 10 words sorted by ambient happiness.  Ambient happiness is calculated using word frequencies from September 2008 through November 2015.  Non-English words and words with frequencies under 1000 are removed, leaving 9789 remaining in our ambient dataset.}
\label{sort_amb}
\end{table}

\begin{table}[h]
\resizebox{\columnwidth}{!}{%
\begin{tabular}{lcccclccc}
\\\hline\noalign{\smallskip}
\multicolumn{4}{ c } {\textbf{Top 10}} & \hspace{3mm} & \multicolumn{4}{ c } {\textbf{Bottom 10}}\\\hline\noalign{\smallskip}
Rank & Word & Ambient & labMT && Rank & Word & Ambient & labMT \\\noalign{\smallskip}\hline\noalign{\smallskip}
1. & birthday & 6.82 & 7.78 && 9780. & seas & 4.80 & 6.84\\
2. & b-day & 6.71 & 7.68 &&  9781. & civilians & 4.86 & 5.84\\
3. & merry & 6.90 & 7.56 &&  9782. & defendants & 4.87& 4.26\\
4. & prizes & 6.73 & 7.20 && 9783. & prosecutors & 4.87 & 4.20\\
5. & giveaway & 6.62 & 6.38 && 9784. & welt & 3.77 & 4.04\\
6. & collected & 7.21 & 5.96 && 9785. & pleaded & 4.78 & 3.84\\
7. & entered & 6.65 & 5.82 && 9786. & sentenced & 4.71 & 3.70\\
8. & shipping & 6.61 & 5.46 && 9787. & indicted & 4.81 & 3.60\\
9. & 2-bath & 6.69 & 5.28 && 9788. & suspects & 4.86 & 3.60\\
10. & iya & 6.79 & 4.94 && 9789. & suspected & 4.81 & 3.52 \\\noalign{\smallskip}\hline
\end{tabular}
}
\caption{The top 10 and bottom 10 words according to ambient happiness, sorted by labMT score.}
\label{sort_lab}
\end{table}

\begin{table}[h]
\resizebox{\columnwidth}{!}{%
\begin{tabular}{lcccclccc}
\\\hline\noalign{\smallskip}
\multicolumn{4}{ c } {\textbf{Top 10}} & \hspace{3mm} & \multicolumn{4}{ c } {\textbf{Bottom 10}}\\\hline\noalign{\smallskip}
Rank & Word & Ambient & labMT && Rank & Word & Ambient & labMT \\\noalign{\smallskip}\hline\noalign{\smallskip}
1. & laughter & 5.96 & 8.50 && 9780. & died & 5.76 & 1.56\\
2. & happiness & 6.11 & 8.44 &&  9781. & kill & 5.71 & 1.56\\
3. & love & 6.17 & 8.42 &&  9782. & killed & 5.56 & 1.56\\
4. & happy & 6.48 & 8.30 && 9783. & cancer & 5.93 &1.54\\
5. & laughed & 5.87 & 8.26 && 9784. & death & 5.66 & 1.54\\
6. & laugh & 6.01& 8.22 && 9785. & murder & 5.39 &1.48\\
7. & laughing & 5.71 & 8.20 && 9786. & terrorism & 5.16 & 1.48\\
8. & excellent & 6.31 & 8.18 && 9787. & rape & 5.46 & 1.44\\
9. & laughs & 6.06 & 8.18 && 9788. & suicide & 5.49 & 1.30\\
10. & joy & 6.19 & 8.16 && 9789. & terrorist & 5.19 & 1.30 \\\noalign{\smallskip}\hline
\end{tabular}
}
\caption{The top 10 and bottom 10 words according to labMT score.}
\label{lab}
\end{table}

\begin{figure}[h!]
\centerline{\includegraphics[width=0.455\textwidth]{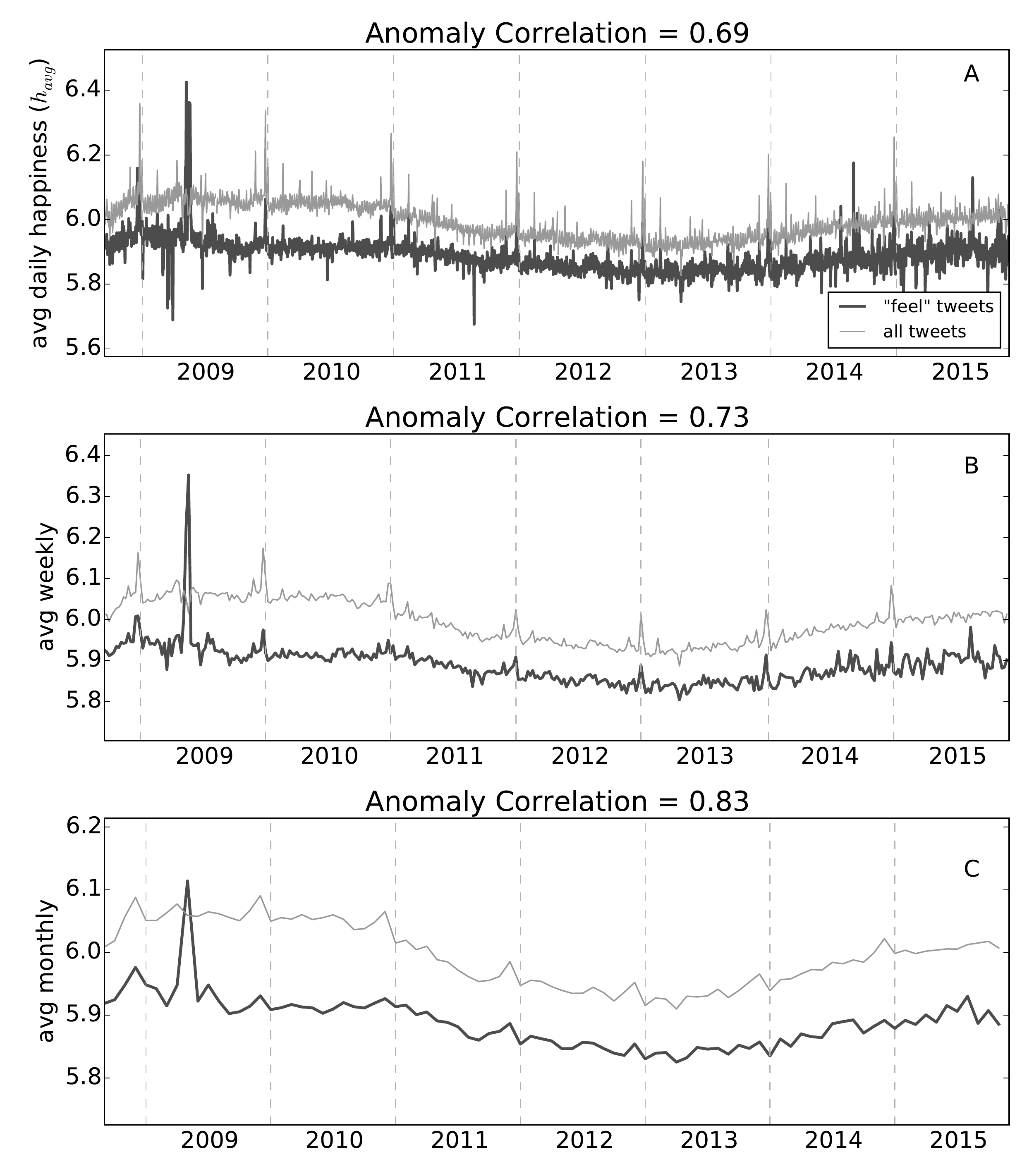}}
\caption{Ambient happiness of ``feel" compared to overall happiness by (A) day,  (B) week, and (C) month (motivated by the work of Kamvar and Harris \cite{kamvar2009we}).  The ambient happiness of the word ``feel" correlates strongly with the average happiness of tweets that do not contain ``feel", and the correlation grows stronger as we decrease the temporal resolution.  This indicates that the shape of overall happiness remains the same whether a user is directly or indirectly expressing an emotion on Twitter.  An interactive version of the overall signal can be found at \url{hedonometer.org}.}
\label{feel}
\end{figure} 

\begin{figure}[h]
\centerline{\includegraphics[width=0.45\textwidth]{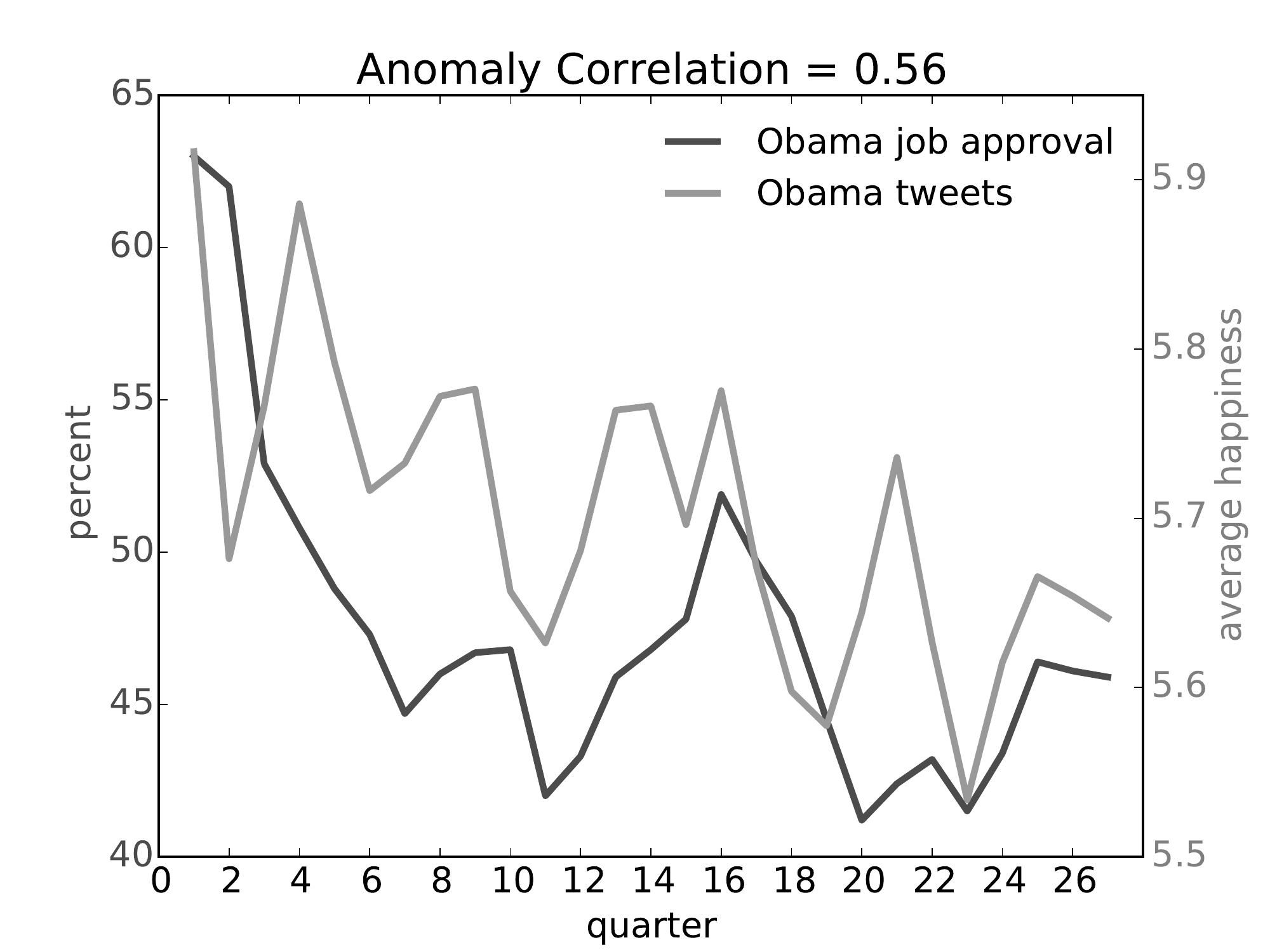}}
\caption{Average quarterly happiness of tweets containing ``Obama" with Obama's quarterly job approval rating from Gallup.  We find a relatively high correlation with solicited polling data.}
\label{obamaquarterA}
\end{figure}  

\begin{figure}[h]
\centerline{\includegraphics[width=0.45\textwidth]{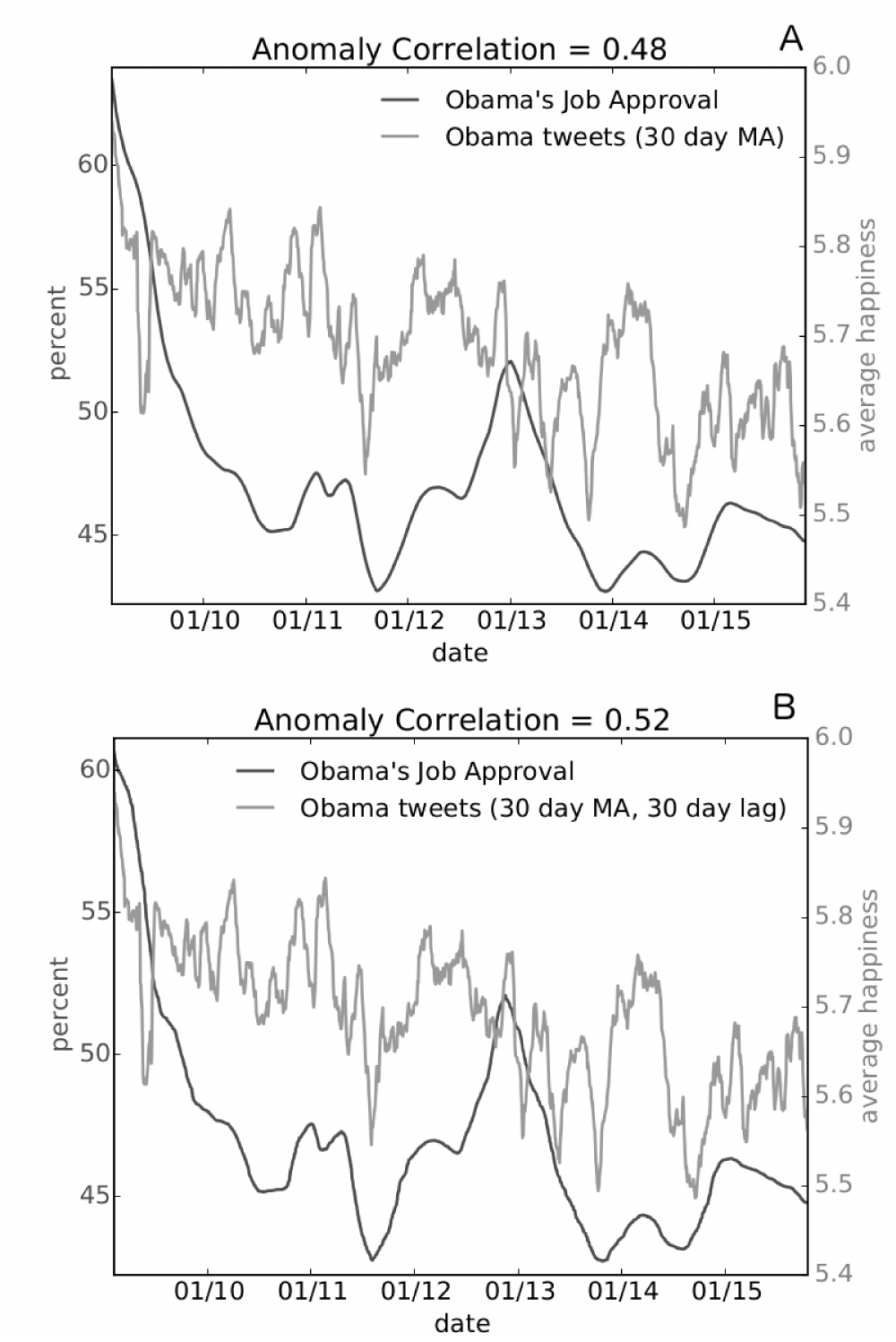}}
\caption{(A) Average daily happiness of tweets containing ``Obama" with Obama's daily job approval rating from Pollster.  (B) 30 day lag.  We find a relatively high correlation with solicited polling data.}
\label{obamadailyA}
\end{figure}  

%

\begin{figure*}[h!]
\centerline{\includegraphics[width=\textwidth]{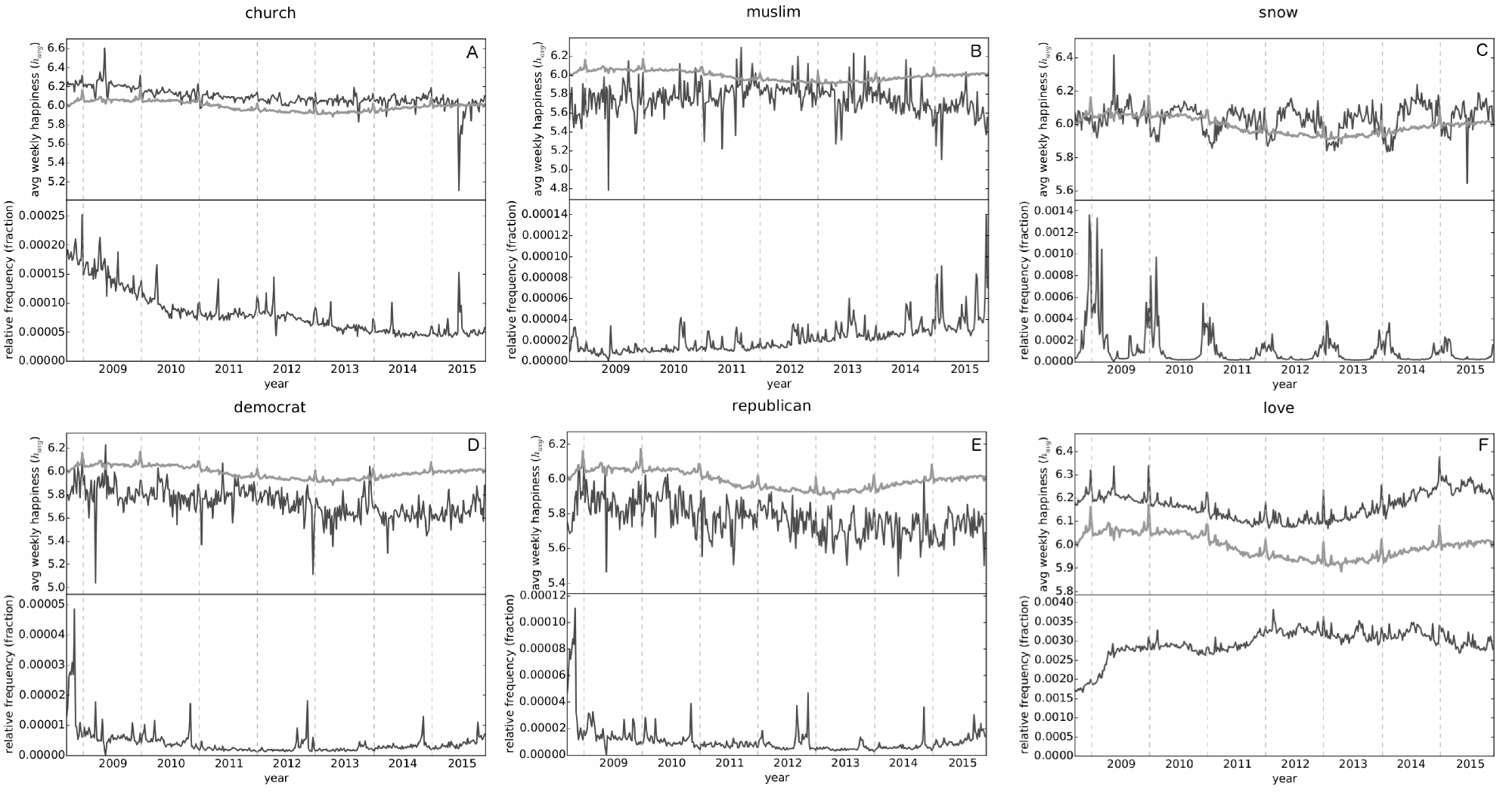}}
\caption{Six examples of weekly ambient happiness time series (top) with the weekly relative frequency for the word (bottom).  Relative frequency is calculated by dividing the total frequency of the word by the total frequency of all words on a given week.  (A) ``church"  (B) ``mulsim" (C) ``snow" (D) ``democrat"  (E) ``republican"  (F) ``love"}
\label{bigfig_week}
\end{figure*}

\begin{figure*}[h!]
\centerline{\includegraphics[width=\textwidth]{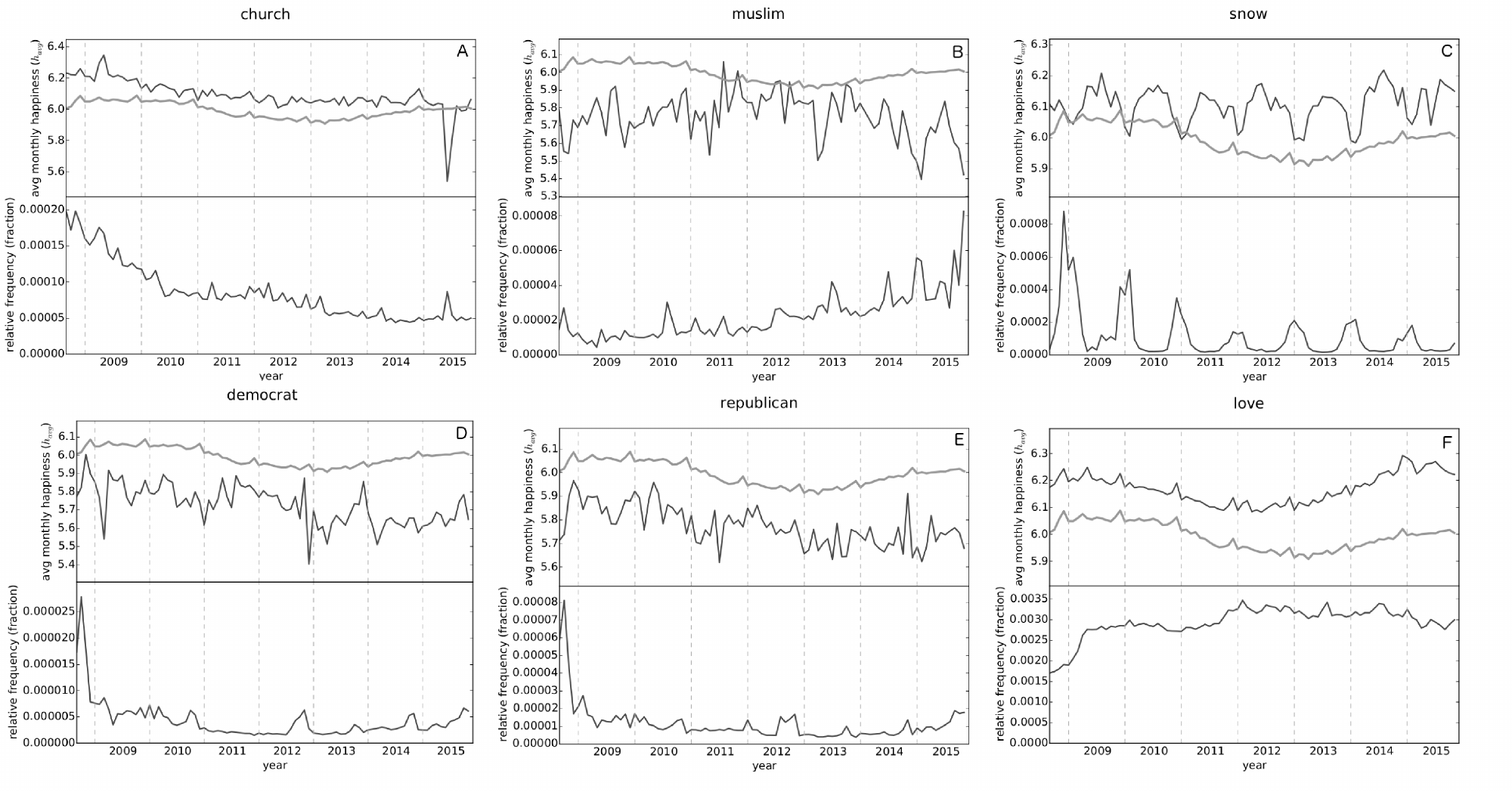}}
\caption{Six examples of monthly ambient happiness time series (top) with the monthly relative frequency for the word (bottom).  Relative frequency is calculated by dividing the total frequency of the word by the total frequency of all words on a given month.  (A) ``church"  (B) ``mulsim" (C) ``snow" (D) ``democrat"  (E) ``republican"  (F) ``love"}
\label{bigfig_month}
\end{figure*}

\section{Gallup Yearly Polling}

\begin{figure}[b!]
\includegraphics[width=\textwidth]{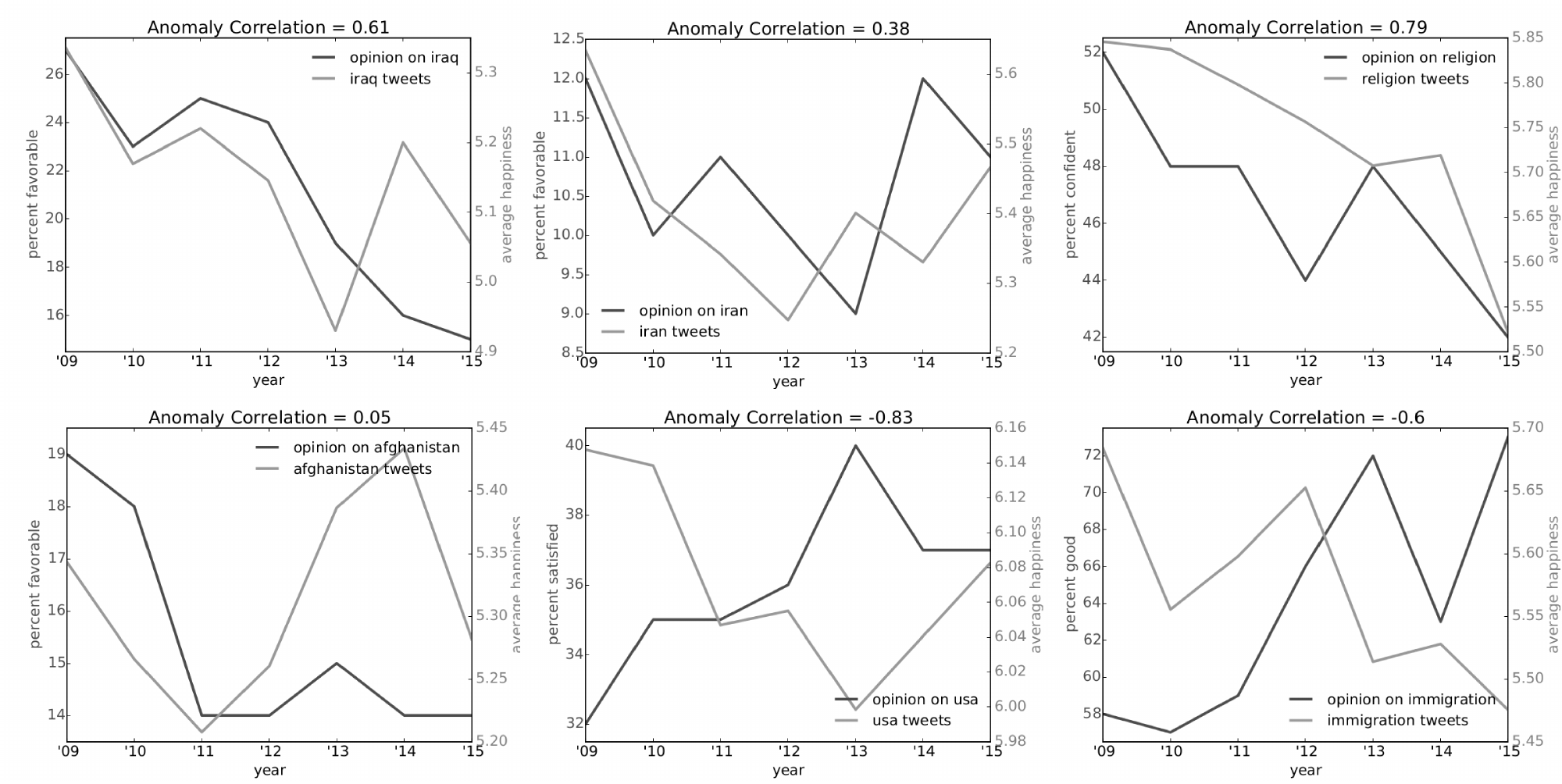}
\caption{Correlations between average ambient happiness and opinion polls on various global subjects.  We obtain varying levels of correlation between the topics due the limited availability of traditional polling data.  For example, Twitter sentiment tracks public opinion surrounding Iraq and religion quite well, but performs poorly on Afghanistan.  The specific questions can be found in Table~\ref{questions}.}
\label{good}
\end{figure}

Gallup trends provide yearly polling data on many topics without a subscription.  The Gallup survey questions can be found in Table~\ref{questions}.  These polls, however, take place only once a year in the same month over several days.  This presents a challenge as to the amount of Twitter data we should include in our correlations, as opinions may change.  For each Gallup datapoint, we use the current year's worth of tweets from 2009 through 2015 for various subjects of national or global interest.  Fig.~\ref{good} shows several topics that correlate quite well with ambient happiness on Twitter.  We find that the favorability of two major countries, Iran and Iraq, has a positive correlation with the ambient happiness of ``Iran" and ``Iraq".  We also find that the United States opinion on religion has a strong positive correlation with yearly ambient happiness of ``religion".  Other topics, including the United States opinion on Afghanistan and immigration show no significant correlation to Twitter data.  There is a strong negative correlation between the satisfaction of the position of the United States in the world, indicating there may be some sarcasm associated with ``usa" tweets.  


\begin{table}[h!]
\resizebox{\columnwidth}{!}{%
\begin{tabular}{cLcc}\\\hline\noalign{\smallskip}
Topic & Survey Question & Frequency & Source\\\noalign{\smallskip}\hline\noalign{\smallskip}
Iraq & What is your overall opinion of Iraq?  Is it very favorable, mostly unfavorable, mostly unfavorable, or very unfavorable? & Yearly & Gallup \\&&&\\
Iran & What is your overall opinion of Iran?  Is it very favorable, mostly unfavorable, mostly unfavorable, or very unfavorable? & Yearly & Gallup\\&&&\\
Afghanistan & What is your overall opinion of Afghanistan?  Is it very favorable, mostly unfavorable, mostly unfavorable, or very unfavorable? & Yearly & Gallup\\&&&\\
USA & On the whole, would you say you are satisfied or dissatisfied with the position of the United States int he world today? & Yearly & Gallup\\&&&\\
Religion & Please tell me how much confidence you , yourself, have in the church or organized religion --  a great deal, quite a lot, some, or  very little? & Yearly & Gallup\\&&&\\
Immigration & On the whole, do you think immigration is a good thing or a bad thing for this country today? & Yearly & Gallup\\&&&\\
Obama & Do you approve of disapprove of the way Barak Obama is handling his job as president? & Quarterly & Gallup\\&&&\\
Obama & Average of latest opinion polls on Obama's job approval & Daily & Pollster \\\noalign{\smallskip}\hline
\end{tabular}
}
\caption{Survey questions for polling data from various resources used in our analysis.}
\label{questions}
\end{table} 

\clearpage

\end{document}